\documentclass[a4paper,11pt]{article}
\pdfoutput=1 

\usepackage{jcappub} 

\usepackage[T1]{fontenc} 

\usepackage{url}
\usepackage{color}
\usepackage{comment}
\usepackage{mathrsfs}
\usepackage{hyperref}
\usepackage{braket}
\usepackage[dvipsnames]{xcolor}
\usepackage{natbib}
\usepackage{physics}
\usepackage{calc}
\usepackage{amsmath, amssymb}

\usepackage{here}
\usepackage{siunitx}
\usepackage{float}
\usepackage{url}
\usepackage{latexsym}
\usepackage{listings, jvlisting}
\usepackage{ascmac}
\usepackage{mathcomp}

\usepackage{minitoc}
\usepackage{amsmath}
\usepackage{amssymb}

\newcommand{\bx}{{\mathbf{x}}}

\newcommand{\bk}{{\mathbf{k}}}

\newcommand{\bq}{{\mathbf{q}}}
\newcommand{\br}{{\mathbf{r}}}

\newcommand{\bs}{{\mathbf{s}}}

\newcommand{\bu}{{\mathbf{u}}}

\newcommand{\avrg}[1]{\left\langle #1 \right\rangle}

\newcommand{\tilden}{{\widetilde{\delta n}}}
\newcommand{\tbarn}{\tilde{\bar{n}}}
\newcommand{\tdeltab}{\tilde{\Delta}_b}

\definecolor{olivegreen}{rgb}{0,0.6,0}
\definecolor{mycyan}{rgb}{0.13, 0.62, 0.8}
\definecolor{deepcarminepink}{rgb}{0.94, 0.19, 0.22}

\title{Modeling survey-window
and integral-constraint effects 
on PNG in the galaxy power spectrum with 
light-cone mocks
}

\author[a,b]{Shintaro~Nakano,}
\author[b,c]{Masahiro~Takada,}
\author[d]{Toshiki~Kurita,}
\author[e]{Ryo~Terasawa}
\affiliation[a]{
Department of Physics, Graduate School of Science,\\
 The University of Tokyo, 7-3-1 Hongo, Bunkyo-ku, Tokyo 113-0033, Japan}
\affiliation[b]{
Kavli Institute for the Physics and Mathematics of the Universe (WPI),\\
 The University of Tokyo Institutes for Advanced Study (UTIAS),\\
 The University of Tokyo, Chiba 277-8583, Japan}%
\affiliation[c]{
Center for Data-Driven Discovery (CD3), Kavli IPMU (WPI), UTIAS, The University of
Tokyo, Kashiwa, Chiba 277-8583, Japan}
\affiliation[d]{
Max-Planck-Institut für Astrophysik, Karl-Schwarzschild-Str. 1, 85748 Garching, Germany
}
\affiliation[e]{Center for Frontier Science, Chiba University, 1-33 Yayoi-cho, Inage-ku, Chiba 263-8522, Japan}
\emailAdd{shintaro.nakano@ipmu.jp}

\date{\today}

\abstract{
We develop an analysis framework based on {\em light-cone} galaxy mock catalogs constructed from linear theory simulations in order to accurately model the impact of primordial non-Gaussianity (PNG) 
on galaxy power spectrum on large scales. 
These linear light-cone catalogs properly incorporate a variety of observational and cosmological effects, including the survey window function,
the redshift evolution of matter and galaxy density fields, and 
redshift-space distortions (RSD).
When estimating the multipole moments of the power spectrum from each light-cone mock, we employ the same estimator as used in actual analyses, thereby properly accounting for the effects of the discrete Fourier transform, the line-of-sight dependence of the fields,  
and wide-angle RSD effects.
Using light-cone mock catalogs that mimic the BOSS survey, we demonstrate that long-wavelength modes comparable to or larger than the survey window scale, namely super-survey modes, 
have a significant impact on the integral constraint (IC) in power spectrum measurements. In particular, we show that the analytical treatment of the survey-window convolution and
IC,
which has been
commonly used in previous studies, begins to lose accuracy on scales of 
$k\lesssim k_{\rm eq}$ (the matter-radiation equality scale), and becomes invalid in the presence of PNG. 
The method developed in this work enables unbiased searches for PNG using the galaxy power spectrum on long-wavelength scales probed by ongoing and future wide-area galaxy surveys.
}

\begin{document}
\maketitle
\flushbottom

\section{Introduction}
\label{sec:introduction}

Wide-area spectroscopic galaxy surveys are becoming a powerful probe of cosmology, 
complementing observations of cosmic microwave background
 (CMB) anisotropies~\cite{SDSS:2005xqv,2dFGRS:2005yhx,2022arXiv221208488I,2022PhRvD.105h3517K,2022JCAP...02..008C,2025JCAP...02..021A,2025JCAP...07..028A}.
In contrast to the CMB, which provides essentially two-dimensional information on the sky, wide-area spectroscopic surveys probe the underlying dark matter distribution and metric perturbations, such as gravitational potential field,  
through the {\it three-dimensional} distribution of galaxies
on large scales~\cite{1980lssu.book.....P} \citep[also see Fig.~5 in Ref.][]{2009PhRvD..79h3530T}. 
Moreover, on sufficiently large scales where density fluctuations remain in the linear regime, the time evolution of perturbations can be accurately predicted within 
the framework of general relativity~\citep[e.g.,][]{2020moco.book.....D,2022cosm.book.....B,2018PhR...733....1D}. 
This enables us to 
infer the properties of primordial fluctuations encoded in the initial conditions of the Universe, including those generated during inflation, from measurements of the large-scale galaxy distribution~\citep{1986PThPh..76.1036S,1988ZhETF..94....1M,1984PThPS..78....1K}.

Examples of cosmologically useful signals encoded in the large-scale distribution of galaxies include the matter–radiation equality scale~\cite{2021PhRvD.103b3538P,2025PhRvD.112f3553B}, 
the shape and features of the primordial curvature perturbation power spectrum~\cite{2005PhRvD..71j3515S}, 
the neutrino free-streaming scale~\cite{1998PhRvL..80.5255H,2006PhRvD..73h3520T,2018PhRvD..97l3540S}, dark energy clustering~\cite{2004PhRvD..69h3503B,2006PhRvD..73h3520T}, and primordial non-Gaussianity (hereafter PNG)~\cite{2008PhRvD..77l3514D,2008JCAP...08..031S,2013MNRAS.428.1116R,2019JCAP...09..010C,2023PhRvD.108h3533K,2025JCAP...06..029C,2026arXiv260405213R}.
In particular, {\it local-type} PNG induces
a characteristic scale-dependent enhancement in the galaxy power spectrum on large scales, 
scaling as $1/k^4$ in the limit of small $k$~\citep{2008PhRvD..77l3514D}.
Consequently, the sensitivity to PNG improves as increasingly larger scales become accessible with high statistical significance
in galaxy power spectrum measurements.

However, several sources of uncertainty complicate accurate measurements of the galaxy power spectrum on large scales.
First, the survey window function, arising from a finite survey geometry, induces a non-trivial mixing between different Fourier modes. As a result, the power spectrum measured in a given $k$ bin receive contributions from neighboring Fourier modes 
over a characteristic width set by the survey window~\cite{FKP}.
This window effect is commonly accounted for on the theory side when comparing model predictions with measurements~\cite{Beutler_2014,FKP_P0,2018AJ....156..160H,2017MNRAS.464.3121W,2021JCAP...11..031B,Hand.Li.ea2017jul,2022PhRvD.105l3501K}. 
Specifically, the two-point correlation function of the survey window function is measured
and used to convolve the theoretical power spectrum, thereby incorporating the effect of the survey window into theoretical predictions.
However, this procedure employs a model power spectrum evaluated at a single effective redshift, thereby neglecting the redshift evolution of both the underlying fluctuation fields and the galaxy distribution. 
Furthermore, it relies on an approximate treatment of the line-of-sight direction, assuming a simplified relation between the position-dependent line of sight within the survey volume and the Cartesian coordinate system used for the Fourier transform~\cite{2006PASJ...58...93Y,2015PhRvD..92h3532S,2015MNRAS.453L..11B,2022PhRvD.105l3501K}. 
In addition, practical clustering analyses require the integral constraint (IC) to be properly imposed. 
The IC arises because 
the background galaxy distribution, which is generally redshift dependent, must be estimated 
from the observed galaxy distribution itself and 
the density {\it fluctuation} field is 
subsequently defined
relative to this estimated background~\cite[e.g., see the FKP estimator in Ref.][for the standard approach]{FKP}. 
Since this procedure is inherently imperfect, it can introduce systematic uncertainties~\citep[see Ref.][for a related discussion]{2019JCAP...08..036D}. 
Such uncertainties can have a significant impact on power spectrum measurements on large scales, particularly in the presence of PNG, 
where the signal is strongest on the largest scales and the effects of the IC have not yet been fully quantified~\cite[see][for a related discussion]{2025arXiv250112661T}.

Motivated by the considerations above, the primary goal of this paper is to investigate in detail the impact of the survey-window convolution and the IC on power spectrum measurements.
In particular, we focus on the role of {\it super-survey} or {\it super-sample} modes~\cite{2013PhRvD..87l3504T,2014PhRvD..89h3519L}, 
namely Fourier modes with wavelengths comparable to or larger than the survey size, and examine how they affect the survey-window convolution and the IC in galaxy power spectrum measurements, especially in the presence of PNG.
To this end, 
we propose the use of light-cone galaxy mock catalogs constructed from linear theory 
simulations as a useful tool for investigating these effects. 
Our light-cone method incorporates 
a wide range of relevant observational and cosmological effects. These include the survey window convolution for a BOSS-like 
survey~\cite{2016AJ....151...44D} as a working example, the discrete Fourier transform, the IC, 
the galaxy and peculiar velocity fields predicted by the $\Lambda$CDM model, 
the large-scale enhancement of galaxy clustering induced by PNG, the redshift distribution (selection function) of galaxies, 
RSD effects beyond the plane-parallel approximation
~\cite{1987MNRAS.227....1K,Hamilton_1998}, and the redshift evolution of all these quantities. 
As a result, we show that the conventional approach described above fails to accurately model the observed galaxy power spectrum on large scales, and that this limitation becomes particularly severe in the presence of PNG.
The methodology developed in this work will be useful for the analysis of ongoing and future galaxy surveys, including Subaru PFS~\cite{2014PASJ...66R...1T}\footnote{\url{https://pfs.ipmu.jp/intro.html}},
DESI\footnote{\url{https://www.desi.lbl.gov}},
HETDEx\footnote{\url{https://hetdex.org}},
SPHEREx\footnote{\url{https://science.nasa.gov/mission/spherex/}},
Euclid\footnote{\url{https://www.esa.int/Science_Exploration/Space_Science/Euclid}}, and
the Nancy Grace Roman Space Telescope\footnote{\url{https://science.nasa.gov/mission/roman-space-telescope/}}.

The structure of this paper is as follows. 
In Section~\ref{sec:galaxy_density_field}, we review the definition of the galaxy density fluctuation field. In particular, we describe the procedure used in actual clustering analyses to estimate the background galaxy distribution from the observed galaxy distribution and to define the galaxy density fluctuation field relative to it.
In Section~\ref{sec:integral_constraint},
we define the power spectrum estimator, 
paying particular attention to the implementation of the survey-window convolution and the IC. 
In Section~\ref{sec:simulations}, 
we describe our method for constructing light-cone galaxy mock catalogs for a BOSS-like survey based on linear theory.
In Section~\ref{sec:results}, we present the main results of this paper. Finally, Section~\ref{sec:conclusion} is devoted to 
a discussion of the results and our conclusions.

Throughout this paper, we use the following abbreviated notations: 
\begin{align}
\int_{\bx }\equiv \int\!\mathrm{d}\bx, \hspace{1em}
\int_{\bk}\equiv \int\!\frac{\mathrm{d}\bk}{(2\pi)^3}.
\end{align}
We also adopt the following notations for the Fourier and inverse Fourier transforms:
\begin{align}
f(\bk)\equiv\int_{\bx}f(\bx)e^{-i\bk\cdot\bx},\hspace{1em}
f(\bx)\equiv\int_{\bk}f(\bk)e^{i\bk\cdot\bx}~ .
\end{align}

\section{Galaxy density fluctuation field and FKP estimator}
\label{sec:galaxy_density_field}

Using a survey window function, the observed density field of galaxies can be expressed as
\begin{align}
\tilde{n}(\bx)=W(\bx)n(\bx)=W(\bx)\bar{n}(z)\left[1+\delta(\bx)\right],
\end{align}
where $n(\bx)$ 
is the underlying number density field, $\delta(\bx)$ is the underlying density fluctuation field, 
and $\bar{n}(z)$ is the {\it true} mean background density and depends only on redshift. 
The window function is defined as
$W(\bx)=1$ if $\bx$ lies within the survey region, and $W(\bx)=0$ otherwise. 
It is straightforward to
further include 
weighting in the observed field, but for simplicity we do not consider it.
Hereafter we use the tilde symbol $\tilde{~}$  to denote fields and quantities defined within the survey region, i.e. 
the observed fields and quantities, 
in contrast to the underlying true fields and quantities.

The true $\bar{n}(z)$ is, of course, not available for any sample. 
It must be estimated from the observed data themselves.
Due to a radial (or redshift-dependent) selection function,
the mean galaxy density generally needs to be estimated from the volume average within each redshift shell:
\begin{align}
\tilde{\bar{n}}(z_i)
\equiv
\frac{1}{V(z_i)}\int_{z_i}\!\frac{\mathrm{d}z}{H(z)}r(z)^2
\int\!\mathrm{d}^2\boldmath{\Omega}_{\hat{\bx}}~
\tilde{n}(\bx)=\bar{n}(z_i)\left[1+\tilde{\Delta}_b(z_i)\right], 
\label{eq:dndz_def}
\end{align}
where we have assumed that an observer is at the coordinate origin,
$H(z)$ denotes the Hubble expansion rate at redshift $z$, $r(z)$ denotes the comoving angular diameter distance to $z$, 
$V(z_i)\equiv\int_{z_i}\!(\mathrm{d}z/H(z))~r(z)^2\int\!\mathrm{d}^2\boldmath{\Omega}~
W(\bx)$, 
$z_i$ denotes the central redshift of the $i$-th redshift shell, and the integration range of $z$ is 
$[z_i-\Delta z/2,z_i+\Delta z/2]$, with width $\Delta z$. $\tilde{\Delta}_b(z_i)$ 
is the volume-averaged density fluctuation within the $i$-th redshift shell centered at $z_i$, defined by
\begin{align}
\tilde{\Delta}_b(z_i)\equiv 
\frac{1}{\bar{n}(z_i)V(z_i)}
\int_{z_i}\!\frac{\mathrm{d}z}{H(z)}r(z)^2
\int\!\mathrm{d}^2\boldmath{\Omega}_{\hat{\bx}}~W(\bx)\bar{n}(z)\delta(\bx)~,
\label{eq:Deltab_def}
\end{align}
where we have assumed the redshift shell is sufficiently thin. 
Note that $\tdeltab(z)$ is an unknown function, and therefore introduces an uncertainty. 
A typical amplitude of $\tdeltab(z_i)$ can be roughly estimated from its variance: 
\begin{align}
\avrg{\left(\tdeltab{}(z_i)\right)^2}&=
\int_{\bk}P(k;z_i)|W_\perp(k_\perp;z_i)|^2|W(k_\parallel;z_i)|^2\nonumber\\
&=\int_0^\infty\!\frac{\mathrm{d}\ln k_\perp}{(2\pi)^2}k_\perp^2|W_\perp(k_\perp;z_i)|^2
\int_0^{\infty}\!\frac{\mathrm{d}\ln k_\parallel}{2\pi}
k_\parallel 
|W(k_\parallel;z_i)|^2
P\!\left(\sqrt{k_\perp^2+k_\parallel^2}\right),\nonumber\\
&\simeq \int_0^{k_{\perp}^W}\!\mathrm{d}\ln k_\perp
\left.\int_0^{k_{\parallel}^W}\!\mathrm{d}\ln 
k_\parallel~
\frac{k_\perp^2k_\parallel}{(2\pi)^3}P\!\left(\sqrt{k_\perp^2+k_\parallel^2}\right)\right|_{k_\perp^W\ll k_\parallel^W}, 
\label{eq:vari_deltab_zi}
\end{align}
where we have approximated the window function of the $z_i$-shell by a slab; we denote by $W_\perp(k_\perp; z_i)$ the window function in the 
direction perpendicular to the line-of-sight direction at redshift $z_i$, 
and by $W_\parallel(k_\parallel;z_i)$ the window function corresponding to the 
redshift width. 
In the third equality, we restrict the integration range to 
$[0,k^W]$, 
using the fact that the window function has significant support only on scales up to the survey-window size. Here $k^W_{\parallel}$ and $k^W_\perp$ denote the wavenumbers corresponding to the survey-window scale in the directions parallel and perpendicular 
to the line of sight, respectively. For a slab window geometry, $k_\perp^W\ll k_\parallel^W$. 

\begin{figure}
\centering
\includegraphics[width=0.6\textwidth]{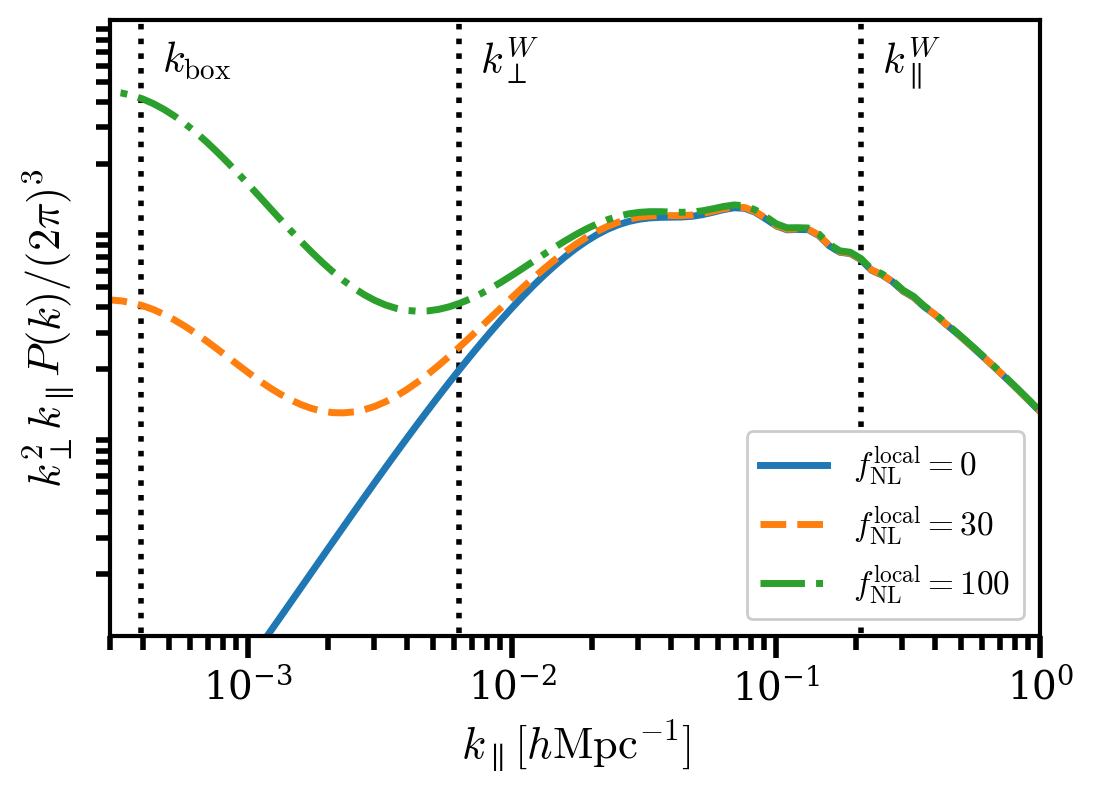}
\caption{The $k$-dependence of the integrand of the variance of the 
volume-averaged density fluctuation within a redshift shell,
$\avrg{(\tdeltab(z_i))^2}$, 
as given by Eq.~(\ref{eq:vari_deltab_zi}). 
Here we approximate the redshift shell as a slab, where 
the characteristic  scale in the R.A. and Dec. directions is $k_\perp$ in units of $h{\rm Mpc}^{-1}$, while the scale corresponding to the redshift width is $k_\parallel$, with $k_\perp \ll k_\parallel$. In our study, we use light-cone simulations constructed from linear-theory simulations with a box size of 
$16~h^{-1}{\rm Gpc}$, whose fundamental Fourier mode is denoted by $k_{\rm box}$.
We also denote by $k^W_\perp$   
  the characteristic Fourier scale of the BOSS survey window in the transverse direction, corresponding to a length scale of 
$O(1)~h^{-1}{\rm Gpc}$ (see below), and by $k_\parallel^W$
the Fourier scale corresponding to the width of a redshift slice with $\Delta z\sim 0.01$. 
The solid, dashed and dot-dashed lines are the predictions for the integrand
for $f_{\rm NL}=0, 30$ and 100, respectively.
Here, we fix $k_\perp=k_{\rm box}$ and plot the integrand as a function of $k_\parallel$, where $k=\sqrt{k_\perp^2+k_\parallel^2}$.
Note that the $y$-axis is shown on a logarithmic scale.
}
\label{fig:vari_deltab_zi}
\end{figure}

Fig.~\ref{fig:vari_deltab_zi} shows the $k$-dependence of the integrand 
of $\avrg{(\tdeltab(z_i))^2}$ for three cases: $f_{\rm NL}=0, 30$ and $100$.
For a BOSS-like survey, $k^W_\perp\sim \mbox{a few}\times O(10^{-3})~h{\rm Mpc}^{-1}$ and $k^W_\parallel\sim \mbox{a few}\times O(10^{-1})~h{\rm Mpc}^{-1}$, because 
the transverse size of the BOSS survey is of order  $ h^{-1}{\rm Gpc}$, whereas $k^W_\parallel$ corresponds to the inverse of the redshift shell width,
$\Delta z/H(z_i)$ with $\Delta z\simeq 0.01$. For the case of $f_{\rm NL}=0$, the contribution from $k_\perp$ is dominated by modes around $k^W_\perp$, and then the contribution from $k_\parallel$ is dominated by modes around the matter-radiation equality scale $(k_{\rm eq})$, corresponding to the peak of the curve shown in the figure. As a result, 
$\avrg{(\tdeltab(z_i))^2}$  receives its dominant contribution from modes around
$k_{\rm eq}$.
For the case of $f_{\rm NL}=30$, modes around both $k\simeq k_{\rm box}$ and $k_\perp^W$
contribute comparably to the $k_\perp$-integral,
where $k_{\rm box}$ is the fundamental mode of simulation box used in this work. The contribution from $k_\parallel$ remains dominated by modes around
$k_{\rm eq}$. 
Therefore, 
$\avrg{(\tdeltab(z_i))^2}$ has the largest contribution at modes around $k_{\rm eq}$.
For the case of $f_{\rm NL}=100$,
 modes around $k_{\rm box}$ have a more significant contribution than 
 modes around $k_{\rm eq}$.
In the real Universe, modes around $k\sim 0$ exist, which would lead to an apparent divergence of $\avrg{(\tdeltab(z_i))^2}$
in the presence of PNG.
However, modes with $k\ll k^W$ are effectively absorbed into the background density and therefore do not affect power-spectrum measurements~\cite{2013PhRvD..87l3504T}.
As a result, 
the IC imposed in power-spectrum measurements removes the effect of this apparent divergence. The measured power spectrum is thus affected not by the $k\sim 0$ modes themselves, but rather by long-wavelength modes with wavenumbers comparable to the characteristic scale of the survey volume, $k^W$, as we will show below.

In addition, the observed density field is affected by shot noise fluctuation, denoted by ${\delta}_{\rm SN}(\bx)$, arising from the finite number of galaxies observed. Taking this contribution into account, the volume-averaged redshift distribution of galaxies 
in each redshift shell is rewritten as
\begin{align}
\tilde{\bar{n}}(z_i)=\bar{n}(z_i)\left[1+\tilde{\Delta}_b(z_i)+\bar{\delta}_{\rm SN}(z_i)\right],
\label{eq:obsdelta_all}
\end{align}
where $\bar{\delta}_{\rm SN}(z_i)$ is the shot noise contribution averaged over the redshift shell, defined in a manner analogous
 to Eq.~(\ref{eq:Deltab_def}).

Therefore, the {\it observed} number density fluctuation field, defined {\it relative} to the estimated background number density,
is given by
\begin{align}
\tilden(\bx)&=W(\bx)\left[{n}(\bx)-\tilde{\bar{n}}(z)
\right]
\nonumber\\
&=W(\bx)\bar{n}(z)\left[\delta(\bx)+\delta_{\rm SN}(\bx)-\tdeltab(z)-\bar{\delta}_{\rm SN}(z)\right]\nonumber\\
&\simeq W(\bx)\tilde{\bar{n}}(z)\left[\delta(\bx)+\delta_{\rm SN}(\bx)-\tilde{\Delta}_b(z)-\bar{\delta}_{\rm SN}(z)\right]~.
\label{eq:deltan_def}
\end{align}
In the third line
we have adopted the linear approximation $|\delta|, |\delta_{\rm SN}|, |\tdeltab|, |\bar{\delta}_{\rm SN}|\ll 1$ and expressed $\tilden(\bx)$ in terms of 
the observed redshift distribution of galaxies, $\tbarn(z)$.
Throughout this paper we adopt the continuous 
density field approximation, $\delta(\bx)$, rather than a discrete particle distribution.

Following the FKP method~\cite{FKP}, we define the galaxy density fluctuation field (hereafter the FKP field)
as
\begin{align}
F(\bx)\equiv \tilden(\bx)=
W(\bx)\tbarn(z)\left[\delta(\bx)+\delta_{\rm SN}(\bx)-\tdeltab(z)-\bar{\delta}_{\rm SN}(z)\right].
\label{eq:F_def}
\end{align}
The FKP field satisfies $\int_{z_i}\mathrm{d}z~ r^2/H(z)\int\!\mathrm{d}\Omega_{\hat{\bx}}~
F(\bx)=0$ over each redshift shell, which is 
sometimes referred to as the radial integral constraint (RIC).
This automatically satisfies $\int_{\bx} F(\bx)=0$, referred as to the global IC (GIC).
Note that the relativistic effects and the magnification bias effects are not included in this work.

\begin{figure}[t]
\centering
\includegraphics[width=0.6\textwidth]{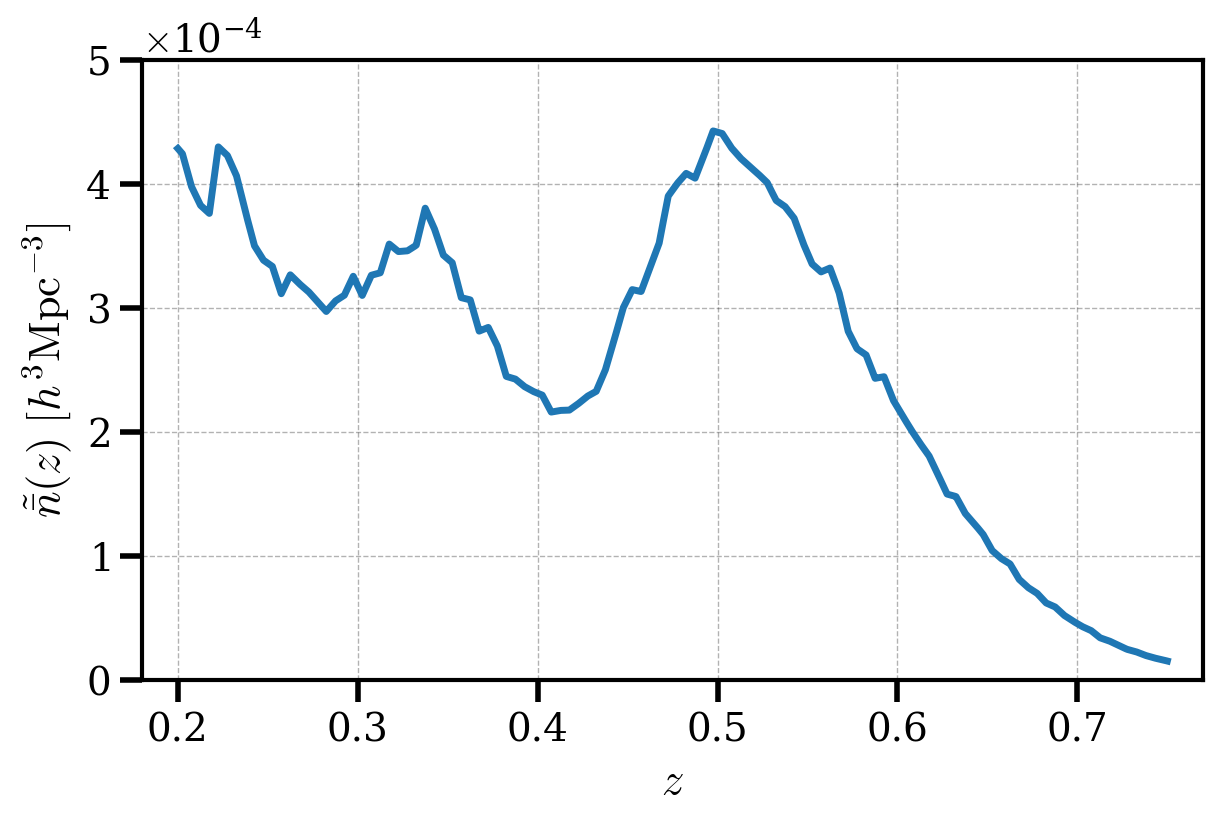}
\caption{The redshift distribution of galaxies, 
$\tbarn(z)$, obtained from the DR12 BOSS catalog (LOWZ plus CMASS samples) in the redshift range $0.2\le z\le 0.75$~\citep{2014MNRAS.441...24A,FKP_P0}. 
The overall shape of $\tbarn$, including the prominent peaks seen at $z\sim 0.35$ and 0.5, is primarily
caused by the selection of LOWZ and CMASS galaxies. 
Apart from the overall shape, it is evident that 
$\tbarn(z)$ is not a smooth function but exhibits fine-scale fluctuations. 
As we discuss in this paper, these fluctuations 
are most likely due to sample variance, namely large-scale density fluctuations within each redshift shell of the light-cone survey volume.
}
\label{fig:dndz}
\end{figure}
\begin{figure}[h]
    \centering
    \includegraphics[width=0.99\textwidth]{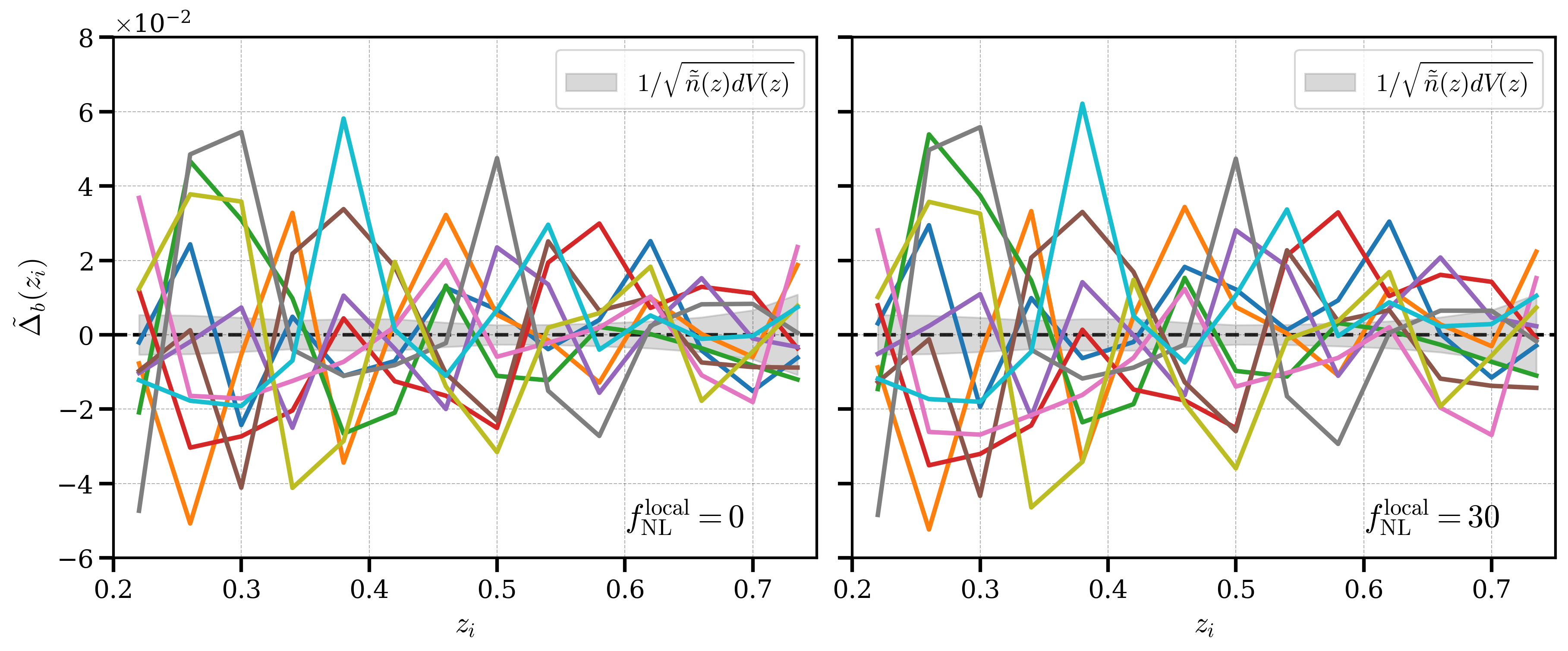}
    \includegraphics[width=0.54\textwidth]{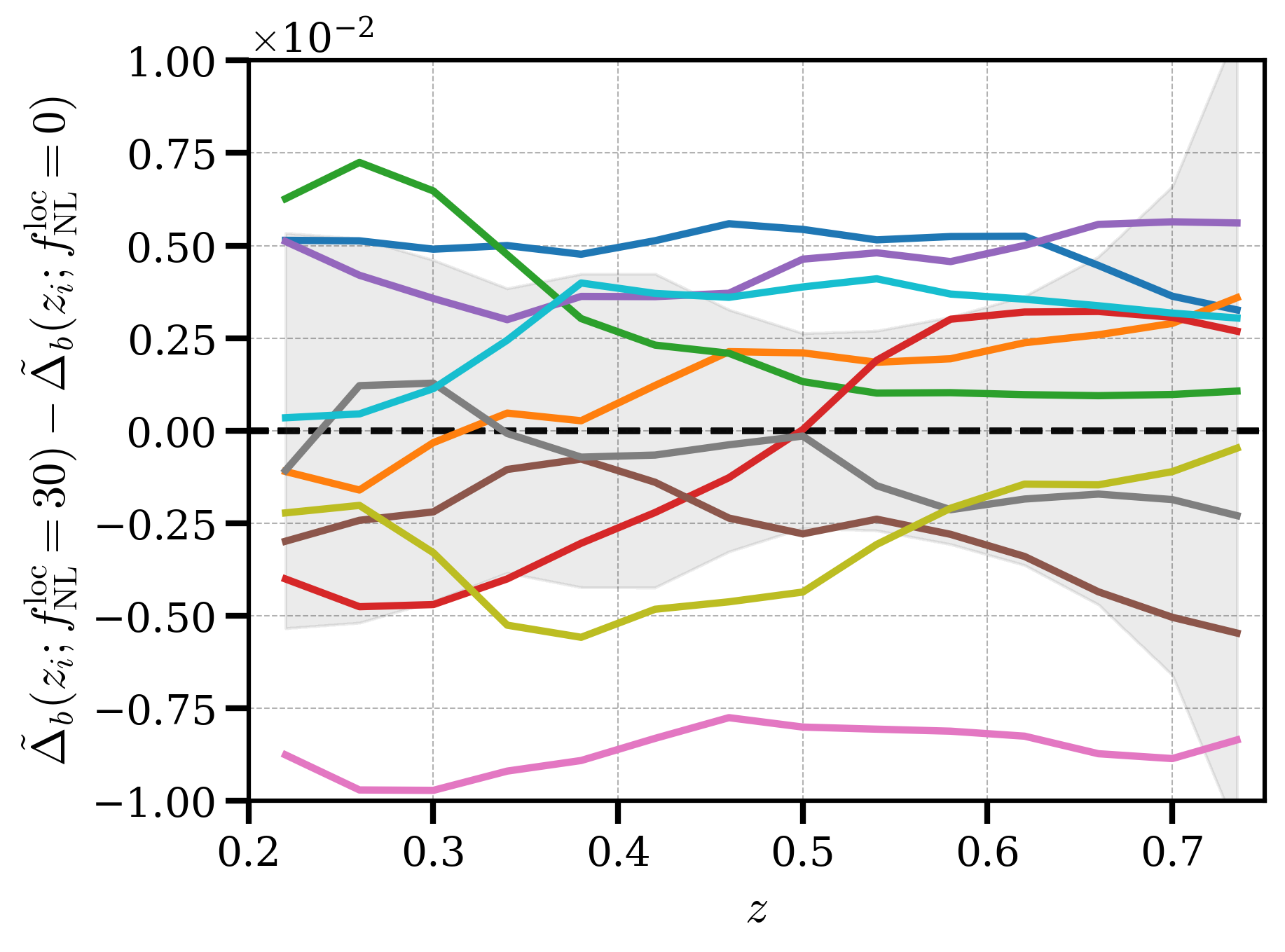}
     \caption{The upper two panels show the volume-averaged density fluctuation,
     $\tdeltab(z_i)$, measured in each redshift shell centered at $z_i$ 
     with a width of $\Delta z=0.04$,  
            for the cases of $f_{\rm NL}=0$ (left panel) and $30$ (right), respectively.
            Each curve shows the result for one of 10 light-cone mock catalogs 
            designed to mimic 
            the BOSS North Galactic Cap (NGC) footprint, covering the  redshift range $0.2\le z\le 0.75$ over an area of
            $\sim 8000~$deg$^2$. Here, we adopt $\Delta z=0.04$ for illustrative purposes, 
            although $\Delta z=0.01$ is used for all other results.
            We use light-cone mocks constructed from linear-field simulations in  
            a $16~h^{-1}{\rm Gpc}$ box
            (see the later sections for details of our methodology). 
            In addition, for the paired realizations with $f_{\rm NL}=0$ and 30 (shown by the same-colored lines),
            we use the same random seeds to generate the linear fields.
            The lower panel shows the difference in $\tdeltab(z_i)$ between the 
            two cases
            for each realization shown in the upper panels.
            The shaded region around zero denotes the $\pm1\sigma$ fluctuation arising from the shot noise assuming the observed BOSS $\tbarn(z)$.
}
\label{fig:dndz_deltab}
\end{figure}
The redshift distribution of galaxies, $\tbarn(z)$, is observable and generally varies with redshift. The random catalog, which determines the background density for a galaxy survey, is constructed using $\tbarn(z)$ to account for the selection effects~\cite[e.g.,][]{2012MNRAS.424..564R}. 
Fig.~\ref{fig:dndz} shows the redshift distribution of galaxies measured from the BOSS survey~\cite{2016AJ....151...44D}, as taken from 
the DR12 data~\footnote{\url{https://dr12.sdss.org/sas/dr12/boss/lss/}}. 
The overall shape of $\tbarn(z)$, including the peaks seen at $z\sim 0.35$ and $z\sim 0.5$, is primarily
caused by the selection of LOWZ and CMASS galaxies through the adopted magnitude and color cuts~\cite{2013AJ....145...10D,2016AJ....151...44D}.
Apart from these overall features, it is also evident that $\tbarn(z)$
is not a smooth function and exhibits fine-scale structures.
In Fig.~\ref{fig:dndz_deltab} we explicitly study how $\tdeltab(z_i)$ and the shot noise give rise to features in the observed $\tbarn(z)$. 
It is clear that $\tdeltab(z_i)$ provides the dominant contribution.
As discussed in Fig.~\ref{fig:vari_deltab_zi}, $\tdeltab(z_i)$ in each redshift slice 
arises primarily from modes
whose wavelengths correspond to modes around $k_{\rm eq}$, 
for both $f_{\rm NL}=0$ and 30 cases. For this reason, $\tdeltab(z_i)$ appear to be similar for the two cases, since the same random seeds are used to generate both light-cone mocks.
However, the difference between the two $\tdeltab(z_i)$, shown in the bottom panel, 
exhibits a coherent fluctuation across the entire redshift range.
This is driven by PNG-induced super-survey modes that coherently span the survey volume.
Thus, it is important to account for the effect 
$\tdeltab(z_i)$ on power spectrum measurements, which is one of the primary goals of this paper.

\section{Analytical approximation for 
survey-window convolution and integral constraint}
\label{sec:integral_constraint}

To facilitate the discussion that follows,
we briefly review the approximation \citep[e.g.,][]{Beutler_2014,FKP_P0} 
for survey-window convolution and the IC, which has been
commonly used in the literature. Before discussing the power spectrum, it is useful to first consider the two-point correlation function. 
An estimator for the two-point correlation function is given by
\begin{align}
\xi^W(\br)&\propto \avrg{F(\bx)F(\bx+\br)}\nonumber\\
&= 
\int_{\bx}
\tbarn(\bx)W(\bx)\tbarn(\bx+\br)W(\bx+\br)\left[\delta(\bx)\delta(\bx+\br)
\vphantom{\tdeltab}
\right.\nonumber\\
&\hspace{2em}\left.-\delta(\bx)\tdeltab(\bx+\br)-\tdeltab(\bx)\delta(\bx+\br)
+\tdeltab(\bx)\tdeltab(\bx+\br)\right],
\label{eq:2pt_def}
\end{align}
where $\tbarn(\bx)$ and 
$\tdeltab(\bx)$ depend only on redshift, i.e. $\tdeltab(z)$ at $z$
corresponding to $|\bx|$ via $r=|\bx|=r(z)$ assuming that an observer is at the coordinate origin. 
The first term on the r.h.s. of Eq.~(\ref{eq:2pt_def}) denotes an estimator for the underlying two-point correlation function,
while the remaining three terms represent the IC contributions arising from $\tdeltab$.
This expression is the most general form and does not rely on any approximation, as also pointed out in Ref.~\citep{2019JCAP...08..036D}. 
Taking the ensemble average of the above estimator and then performing a Fourier transform yields 
a theoretical prediction
for the window-convolved power spectrum, including the IC.
Light-cone mock catalogs enable us to estimate this window-convolved power spectrum while
automatically incorporating all IC contributions.
The approximated treatment of the IC commonly
used in the literature 
is given by
\begin{align}
\xi^W(\br)\propto \int_{\bx}\tbarn(\bx)W(\bx)\tbarn(\bx+\br)W(\bx+\br)
\left[\delta(\bx)\delta(\bx+\br)
-\left(\overline{\tilde{\Delta}}_b\right)^2
\right],
\label{eq:approx}
\end{align} 
where $\overline{\tilde{\Delta}}_b\equiv \sum V(z_i)\tdeltab(z_i)/\sum V(z_i)$
denotes the density contrast averaged over the entire survey volume and is therefore independent of redshift for a given survey realization.
That is, the approximation assumes that the second and third terms in Eq.~(\ref{eq:2pt_def})
are equal to the last term, and further assumes that 
the IC contribution is approximated by the square of the global density contrast,
$\overline{\tilde{\Delta}}_{\rm b}$.

We now turn to the analytical expression for the window-convolved power spectrum including 
the IC. 
First, the window-convolved power spectrum before applying the IC correction is given by
\begin{align}
P^W(\bk; z_{\rm eff})=\int_{\bq }P(\bq; z_{\rm eff})Q(\bk-\bq),
\label{eq:pw_analytic_def}
\end{align}
where $z_{\rm eff}$ is the {\it effective} redshift,
and $P(\bk; z_{\rm eff})$ is the underlying true power spectrum evaluated at $z_{\rm eff}$,
which is generally anisotropic in redshift space.
Thus, the analytical calculation is usually performed at a single effective redshift,
$z_{\rm eff}$.
The function $Q(\bk)$ describes the effect of the survey-window convolution, which corresponds to the Fourier transform 
of $Q(\br) \equiv A^2\int_\bx~\tbarn(\bx)W(\bx)\tbarn(\bx+\br)W(\bx+\br)$ (see below for a normalization constant $A$), which appears in the integrand of Eq.~(\ref{eq:approx}). 
$Q(\br)$ can be estimated from the auto-correlation function of the survey window using a random catalog.
Its multipole moments are 
given by Refs.~\cite{1994MNRAS.267..785C,FKP,Beutler_2014,FKP_P0} as
\begin{align}
Q_\ell(r) &\equiv 
A^2 (2\ell+1) \int \frac{\mathrm{d}^2{\Omega}_{\hat{\br}}}{4\pi} \int_\bx \, 
\tbarn(\bx) W(\bx) 
\tbarn(\bx+\br)W(\bx+\br) 
\mathcal{L}_\ell(\hat{\bx}\cdot\hat{\br}).
\end{align}
where ${\cal L}_\ell(x)$ is the $\ell$-th order Legendre polynomial and 
$\hat{\bx}$ and $\hat{\br}$ are the unit vectors in the directions $\bx$ and $\br$, respectively.
The normalization constant is defined as
\begin{align}
A\equiv \frac{1}{\left[\int_{\bx}W(\bx)^2\tbarn(\bx)^2\right]^{1/2}}.
\label{eq:A_def}
\end{align}
When the IC is imposed, Eq.~(\ref{eq:pw_analytic_def})
is modified \cite{Beutler_2014,FKP_P0,2017MNRAS.464.3121W} as 
\begin{align}
P_\ell^{W,{\rm w/IC}}(k;z_{\rm eff})=P^W_\ell(k; z_{\rm eff})-\frac{Q_\ell(k)}{Q_0(k=0)}
\int_{\bq}P_0(q;z_{\rm eff})
Q_0(\bq).
\label{eq:pw_ell_analytic_def}
\end{align}
The above equation satisfies $P_0^{W,{\rm w/IC}}(k=0;z_{\rm eff}) = 0$. 
This equation corresponds to the Fourier transform of Eq.~(\ref{eq:approx}).
We also note that the second term in the above equation corresponds to Eq.~(\ref{eq:vari_deltab_zi}) under the assumption of no redshift evolution. 
We will use Eq.~(\ref{eq:pw_ell_analytic_def}) below to assess the accuracy of the conventional 
treatment of the IC effect on large scales (small $k$).

\section{Light-cone galaxy mock catalogs based on linear-theory simulations}
\label{sec:simulations}

\subsection{Simulated density fields in real- and redshift-space}
\label{ssec:simulated_fields}
\begin{figure}[t]
\centering
\includegraphics[width=0.99\textwidth]{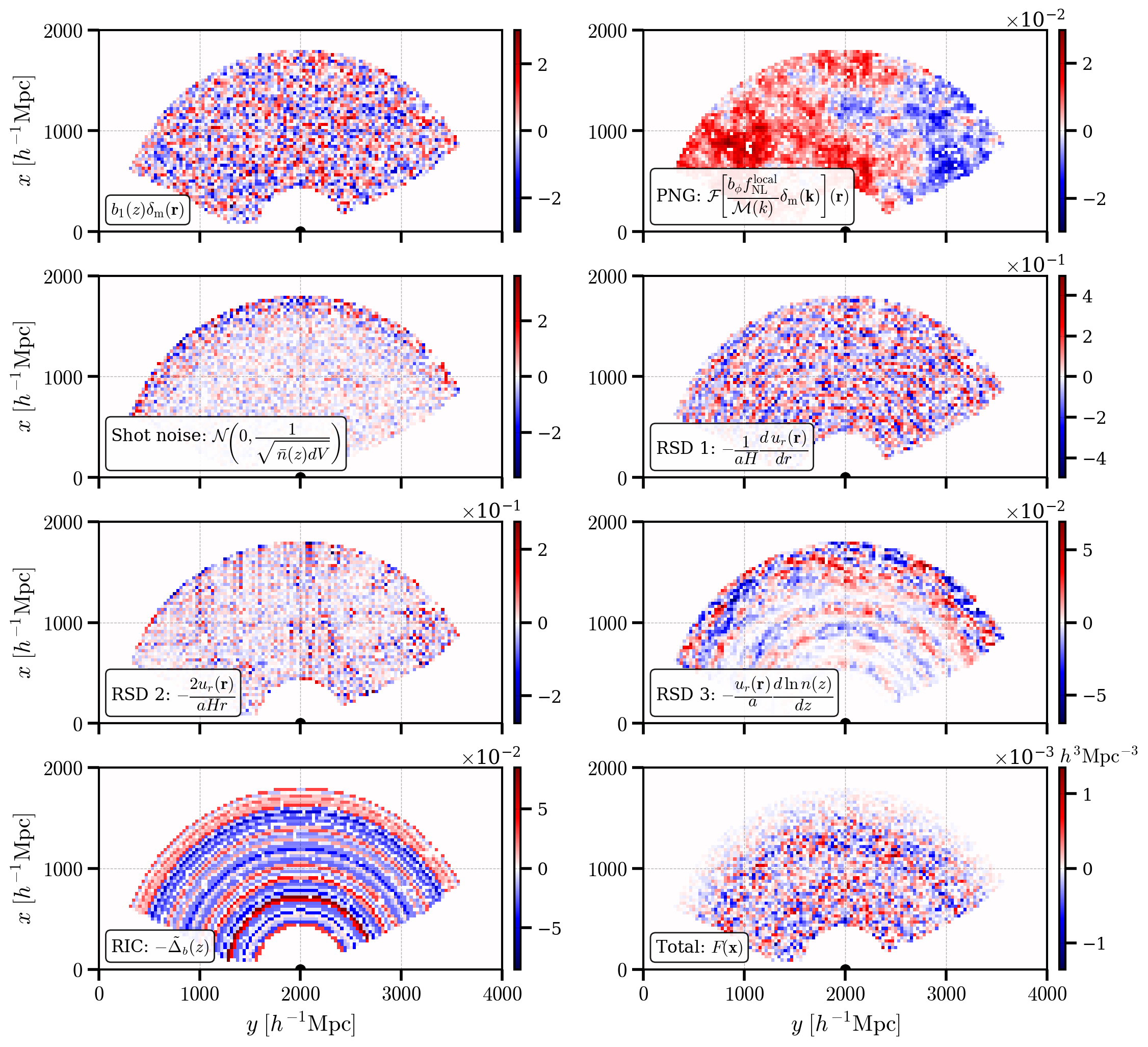}
\caption{Each component of the simulated galaxy density fields in 
a single realization of the linear light-cone mocks. 
Here we consider the BOSS NGC survey footprint (see text for the details).
Note that,
along the direction perpendicular to the page, the field is projected over a thickness corresponding to one grid cell, i.e.,
$16~h^{-1}{\rm Gpc}/512\simeq 31.3~h^{-1}{\rm Mpc}$.
The upper two panels 
show the Gaussian ($f_{\rm NL}=0$) and PNG ($f_{\rm NL}=30$) contributions in the galaxy density fluctuation field, respectively. 
The panel labeled ``Shot noise''
shows the shot noise field.
The three panels labeled ``RSD''
show each contribution of the RSD effect to the redshift-space density fluctuation field, as
given by Eq.~(\ref{eq:RSD_each_term}). The panel labeled ``RIC: $\tdeltab(z)$''
shows the volume-averaged density contrast within each redshift shell centered at $z$,
$\tdeltab(z_i)$ (see Eq.~\ref{eq:Deltab_def}).
The panel labeled ``Total: $F(\bx)$''
shows the redshift-space FKP field including all contributions, in units of $(h^{-1}{\rm Mpc})^{-3}$. 
}
\label{fig:simulated_fields}
\end{figure}

In this paper, we focus on large-scale fluctuation fields and analyze their power spectrum. Since the fluctuations on the scales of interest remain in the linear regime, we employ linear-theory simulations.

We generate each linear-theory
simulation using the fast Fourier transform (FFT) on
$512^3$ grids in a comoving cubic box with a side length of 16~$h^{-1}{\rm Gpc}$. The simulation box is sufficiently large to cover the entire volume of the BOSS North Galactic Cap (NGC) footprint, spanning
the redshift range $0.2\le z\le 0.75$ and covering approximately 8,000~sq. degrees~\cite{2014MNRAS.441...24A,2013AJ....145...10D}. 
For our purpose, we generate fields for the matter density ($\delta_{\rm m}$), galaxy density
($\delta$), and peculiar velocity ($u$) fields,
where the velocity field is needed to simulate RSD~\citep{1987MNRAS.227....1K}. We assume that these fields share the same Fourier phases under the adiabatic initial conditions~\citep{1984PThPS..78....1K} \citep[also see][]{2020moco.book.....D,2022cosm.book.....B}. 
On the other hand, 
local-type PNG 
induces a characteristic scale-dependent enhancement in the galaxy bias on large scales,
as first pointed out in the pioneer 
work~\citep{2008PhRvD..77l3514D}. Based on this consideration, we generate the 
linear fields, $\delta_{\rm m}$, $\delta$
and $u_{\rm m}$, 
 using the FFT method: 
\begin{align}
&\delta_{\rm m}(\bk;{z_{\rm fid}})=
\sqrt{P^L_{\rm m}(k;{z_{\rm fid}})}e^{i\phi_{\bk}}\nonumber\\{}
&\delta(\bk;{z_{\rm fid}})=
b_{\rm g}(k;{z_{\rm fid}})\delta_{\rm m}(\bk; {z_{\rm fid}})\nonumber\\
&
{\bf u}(\bk;{z_{\rm fid}})=
a(z_{\rm fid})H(z_{\rm fid}) f{(z_{\rm fid})} \frac{i{\bf k}}{k^2}\delta_{\rm m}(\bk; {z_{\rm fid}})\ ,
\label{eq:deltam_generation}
\end{align}
with 
\begin{align}
b_{\rm g}(k;{z_{\rm fid}})=b_1(z_{\rm fid})+b_\phi(z_{\rm fid}) f_{\rm NL}{\cal M}^{-1}(k,z_{\rm fid}),
\end{align}
where $P^L_{\rm m}(k;z)$ is the linear matter power spectrum at redshift $z$, $b_{\rm g}$ 
is the linear galaxy bias function, $b_1$ is the linear bias for the Gaussian initial condition, 
and 
$b_\phi$ is the PNG bias parameter. ${\cal M}$ is the transfer function, 
which relates the matter density to the primordial potential in the linear
regime as $\delta_{\rm m}({\bf k};z)={\cal M}(k,z)\Phi({\bf k})$, where 
${\cal M}(k)\equiv (2/3)k^2T(k)D(z)/(\Omega_{\rm m}H_0^2)$ with $T(k)$ and $D(z)$ 
denoting the transfer function and the linear growth factor,
respectively.
To minimize the effect of sample variance, we adopt the fixed amplitude method~\cite{2016MNRAS.462L...1A} as given by
$|\delta_{\rm m}(\bk;{z_{\rm fid}})|=\sqrt{P^L_{\rm m}(k; z_{\rm fid})}$,
where $z_{\rm fid}$ denotes the fiducial redshift at which the input linear power spectrum is computed, and we adopt $z_{\rm fid} =0.47$. 
For each $\bk$
vector, the phase $\phi_{\bk}$
is drawn randomly from a uniform distribution over the interval $[0,2\pi)$.
After performing inverse FFTs on the above fields to obtain the corresponding 
real-space fields, we assign a redshift to each grid cell using the redshift-distance relation, $z=z(r)$ (or $r=r(z)$), 
for an observer at the coordinate origin.
The redshift-dependent fields are then constructed by applying the appropriate redshift-dependent factor; e.g., 
the density field in the absence of PNG is given by 
$\delta(\bx;z)=(b_1(z)D(z)/b_1(z_{\rm fid})D(z_{\rm fid})) \delta(\bx;z_{\rm fid})$.

To compute the linear matter power spectrum and the transfer function, 
we adopt the following cosmological parameters
for a flat $\Lambda$CDM model: 
$\omega_{\rm b}=0.02237$ and $\omega_{\rm c}=0.1200$ for the physical density parameters of baryon and CDM, respectively; $\Omega_{\Lambda}=0.6847$ for the density parameter of $\Lambda$;
and $A_{\rm s}=2.083\times 10^{-9}$ 
and $n_{\rm s}=0.9649$ for the amplitude and spectral tilt of the primordial curvature perturbation power spectrum, respectively. 
Note that the Hubble parameter is $h = 0.6736$, 
and the neutrino mass is $\Sigma m_\nu = 0.06~\mathrm{eV}$ for this model. 
We consider a galaxy sample mimicking the combined sample of LOWZ and CMASS~\cite{2013AJ....145...10D}, and assume that the linear galaxy bias follows 
$b_1(z)=2.040+0.322z$ that is obtained from the linear interpolation of the results 
shown in Table~III of Ref.~\cite{2025PhRvD.111f3548I}. For the PNG bias, we adopt the 
simple relation 
$b_\phi=2(b_1-1)\delta_{\rm cr}$, where $\delta_{\rm cr}=1.686$~\cite{2008PhRvD..77l3514D},
and consider two cases, $f_{\rm NL}=0$ and $30$, to compare the results.

After generating the linear fields in a cubic volume, we embed the BOSS NGC survey volume within it.
We adopt the angular masks publicly available online\footnote{\url{https://dr12.sdss.org/sas/dr12/boss/lss/geometry/}}, where the survey footprint is given by spherical polygons constructed using the MANGLE software~\cite{2008MNRAS.387.1391S}.
We also implement the radial selection function following $\tbarn(z)$ in Fig.~\ref{fig:dndz}.
We retain the simulated linear fields on grids lying inside the BOSS window, while assigning zero values to grids outside the survey window. 
The simulated fields inside the survey window
includes naturally the effects of super survey modes, particularly those arising from the enhanced power of galaxy density fluctuations induced by PNG. When including the RSD effect along the line of sight for an observer at the origin,
we generate the density field in redshift space as follows:
\begin{align}
\delta^S(\bs)\simeq\delta(\bx)
-\frac{1}{aH}\frac{\partial u_r(\bx)}{\partial r}
-\frac{\alpha({r})}{aHr}u_r(\bx)
,
\label{eq:RSD_each_term}
\end{align}
where $r\equiv |\bx|$,
$u_r\equiv \hat{\bx}\cdot\bu$, $\partial u_r/\partial r
\equiv \hat{\bx}\cdot\nabla u_r$
and $\alpha({r})$ is the logarithmic derivative of the galaxy selection function, defined as
$\alpha({r})\equiv \mathrm{d}\ln~r^2\tilde{\bar{n}}(z({r}))/{\mathrm{d}\ln r}$. We can use an FFT method to generate $\delta^S$ from the fields $\delta$ and ${\bf u}$.

When further including the shot noise, we generate a Gaussian random field with the following variance, and then add it to the density fluctuation on each grid: 
\begin{align}
\delta_{\rm SN}({\bx})\sim
N(0,V_{\rm grid}\tilde{\bar{n}}(z)),
\label{eq:def_delta_sn}
\end{align}
where $V_{\rm grid}$ is the volume of a single grid cell. 
Since we are interested only in large-scale modes, the Gaussian approximation, rather than
the Poisson process, is sufficiently accurate.

In summary, we generate the ``observed'' FKP galaxy field
in real- and redshift-space as follows, respectively:
\begin{align}
\tilde{F}(\bx)&=\tbarn(z)\left[\delta(\bx)+\delta_{\rm SN}(\bx)-\tdeltab(z)-\bar{\delta}_{\rm SN}(z)\right],
\nonumber\\
\tilde{F}^S(\bx)&=\tbarn(z)\left[\delta^S(\bx)+\delta_{\rm SN}(\bx)-\tdeltab^S(z)-\bar{\delta}_{\rm SN}(z)\right],
\label{eq:F_def_sim}
\end{align}
where $\tdeltab(z)$, $\tdeltab^S(z)$ and $\bar{\delta}_{\rm SN}(z)$ are the volume-averaged fluctuations 
in each $z$ bin and in each realization 
as discussed in Section~\ref{sec:galaxy_density_field}.
This procedure corresponds to the shuffling method adopted in Ref.~\cite{2026arXiv260405213R} 
though it's based on particle distribution. 
Note that, for redshift-space case, we compute $\tdeltab^S$ from the volume average of the redshift-space 
density field within each redshift shell, and therefore includes the RSD contribution. For notational simplicity, however, 
we omit the superscript ${}^S$ throughout this paper.
We also note that the quantities expressed as a function of $z$ in the above 
equation depend only on redshift and, through $z=z(r)$, is equivalently 
a function only of the distance from the observer. 
For the radial weight function, $\tbarn(z)$, appearing in the above FKP field
and Eq.~(\ref{eq:def_delta_sn}), 
we use the redshift distribution of BOSS galaxies shown in Fig.~\ref{fig:dndz}.

Fig.~\ref{fig:simulated_fields} shows each contribution of the simulated FKP field, for a single realization of the linear light-cone 
mocks, assuming the BOSS NGC survey footprint. In this way, our method enables us to simulate the three-dimensional galaxy distribution on the light cone in a manner that faithfully reproduces actual observations.
Since the FKP field is expressed as a linear sum of its components, 
we can readily evaluate
the auto- and cross-contributions of the different components to the power spectrum.

\subsection{Power spectrum measurements from light-cone mocks}

We treat the light-cone galaxy mocks in the same manner as real galaxy catalogs.
First, for each realization, we estimate the volume-averaged
density fluctuation in each redshift shell, $\tdeltab(z_i)$ 
(and $\bar{\delta}_{\rm SN}(z_i)$ when the shot noise is included),
from the real- or redshift-space density field 
(see Fig.~\ref{fig:simulated_fields}). 
As repeatedly emphasized throughout this paper, the estimated {\it background}
redshift distribution of galaxies
is affected by sample variance.
Accordingly, we define the fluctuation field, i.e. the FKP fields (Eq.~\ref{eq:F_def_sim}), 
on each grid relative to the background density estimated for each realization.

Using the FKP field in each realization, we measure the multipole moments of galaxy power spectrum from the simulated maps using the local-plane-parallel (LPP) estimator \citep{2015PhRvD..92h3532S,2015MNRAS.453L..11B}:
\begin{align}
\hat{P}_{\ell}(k)
&=A^2(2\ell +1)
\int\!\frac{\mathrm{d}^2{\Omega}_{\hat{\bk}}}{4\pi}
F_0(\bk)F_\ell(-\bk)-\hat{P}_{\rm shot},
\label{eq:pkell_estimator}
\end{align}
with 
\begin{align}
F_\ell(\bk)\equiv \frac{4\pi}{2\ell+1}\sum_{m=-\ell}^{\ell}Y_{\ell m}(\hat{\bk})
\int_{\bx} F(\bx)Y^\ast_{\ell m}(\hat{\bx})e^{-i\bk\cdot\bx},
\end{align}
where 
we place an observer at the coordinate origin in the Cartesian coordinate system
used for the Fourier transform, 
$\hat{P}_{\rm shot}$
is the shot noise contribution, 
$Y_{\ell m}$ is the spherical harmonics, 
and $\hat{\bk}$ is the unit vectors in the direction $\bk$. 
The prefactor $A$ is computed from $\tbarn(z)$ for each realization
according to Eq.~(\ref{eq:A_def}). 
The shot noise term is estimated according to the FKP method~\cite{FKP} as
\begin{align}
\hat{P}_{\rm shot}
=A^2 \int_{\bx}W(\bx)^2\tbarn(\bx).
\end{align}
We use an FFT volume of $(4~h^{-1}{\rm Gpc})^3$ with $256^3$ grids, which fully encloses the BOSS NGC survey footprint. 
Unless otherwise stated, we subtract the shot-noise contribution, $\hat{P}_{\rm shot}$, from the measured power spectrum throughout this paper.
Note that the shot noise term is designed to subtract its contribution 
from the measured power spectrum on small scales (i.e., at the limit of large $k$). 
Because the FKP field (Eq.~\ref{eq:F_def}) automatically enforces the IC, subtracting this constant shot-noise term from the measured power spectrum
can lead to negative values of the power spectrum in  the limit of $k\rightarrow 0$. 
This artifact should be kept in mind when 
analyzing actual data~\citep[also see Ref.][for a similar discussion]{2019JCAP...08..036D}.

Using a large number of mock catalogs generated with different random seeds for the phase $\phi_{\bk}$ (see Eq.~\ref{eq:deltam_generation}), 
we obtain well-converged multipole moments from their ensemble average.
This method has several advantages. 
In particular, we can adopt exactly the same FFT setup, including the FFT volume and the number of grids, for power spectrum measurements from both real galaxy catalogs and light-cone mock catalogs.
As a result, the measured power spectrum 
naturally incorporates all relevant observational and numerical effects, including the survey-window convolution, FFT discretization effects
(especially in small-$k$ bins), and the effects associated with the LPP method. For instance, even if the LPP is an approximation and does not capture wide-angle effects~\cite{2015MNRAS.447.1789Y,2018MNRAS.476.4403C}, we can 
nevertheless make a consistent comparison between observations and light-cone mock results, since the power spectrum multipoles are measured in exactly the same manner in both cases.

\section{Results}
\label{sec:results}

\begin{figure}
\centering
\includegraphics[width=0.99\textwidth]{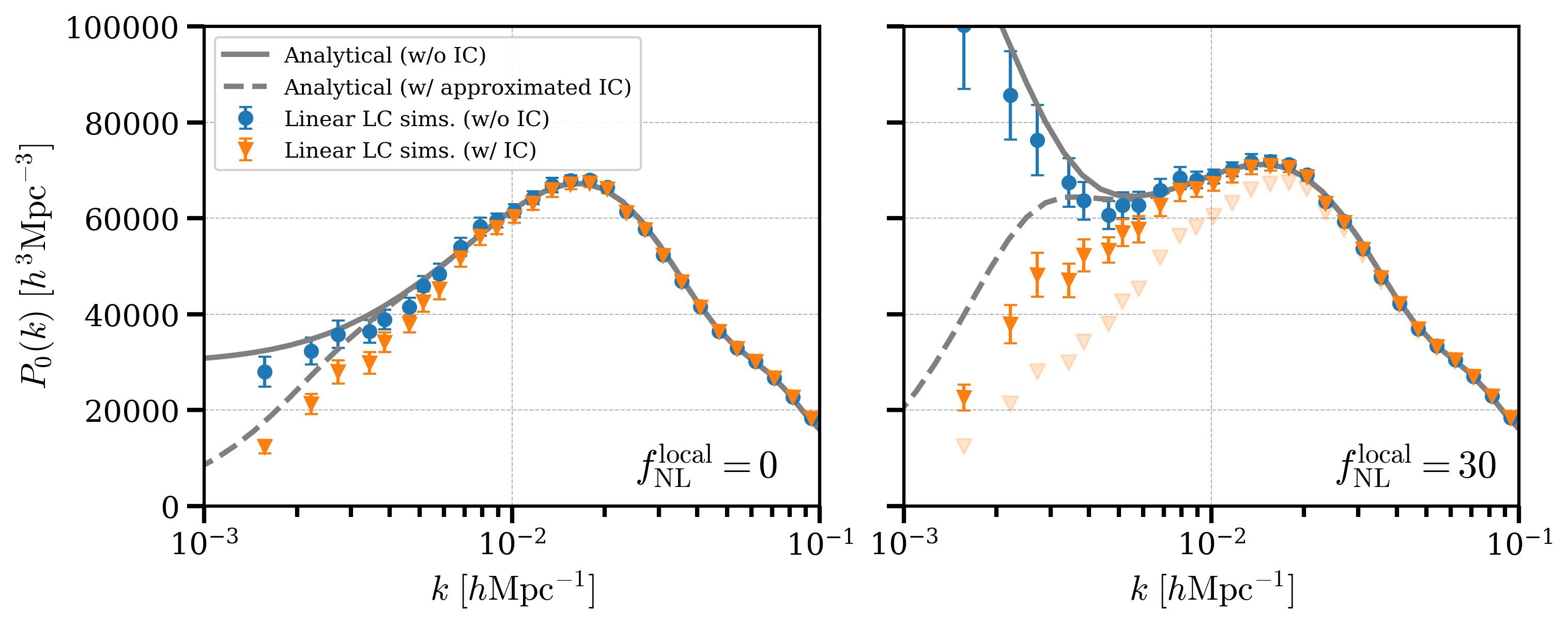}
\caption{The monopole moment of the power spectrum measured from the linear light-cone 
mock catalogs to mimic the BOSS survey footprint in  {\it real} space, i.e., without 
the RSD effect. The blue and orange points show the results obtained without 
and with the integral constraint (IC), respectively. We include other effects, such as 
the redshift evolution of the density fluctuation fields and the galaxy linear bias,
as well as the redshift distribution of galaxies, $\tbarn(z)$
(Fig.~\ref{fig:dndz}), but do not  include the shot noise effect. 
The error bars on each symbol show the $\pm 1\sigma$ 
uncertainty on the mean, estimated from the 50 realizations as $\sigma/\sqrt{50}$,
where $\sigma$ denotes the standard deviation among the 50 realizations.
The solid and dashed 
lines show the analytical predictions for
the window-convolved power spectrum 
with and without the IC, respectively,
assuming 
$z_{\rm eff}=0.462$ as the effective redshift.
This approximation has been
commonly used in the literature (see the text for the details).
For the $f_{\rm NL}=30$ case, we adopt
$k_{\rm min}=2\pi/16~h{\rm Gpc}^{-1}$, 
corresponding to the simulation box size,
as the lower limit of the numerical integration. 
For comparison, the light-color symbols in the right panel shows the result for the $f_{\rm NL}=0$ case.
}
\label{fig:pk0_real}
\end{figure}
\begin{figure}
\centering
\includegraphics[width=0.99\textwidth]{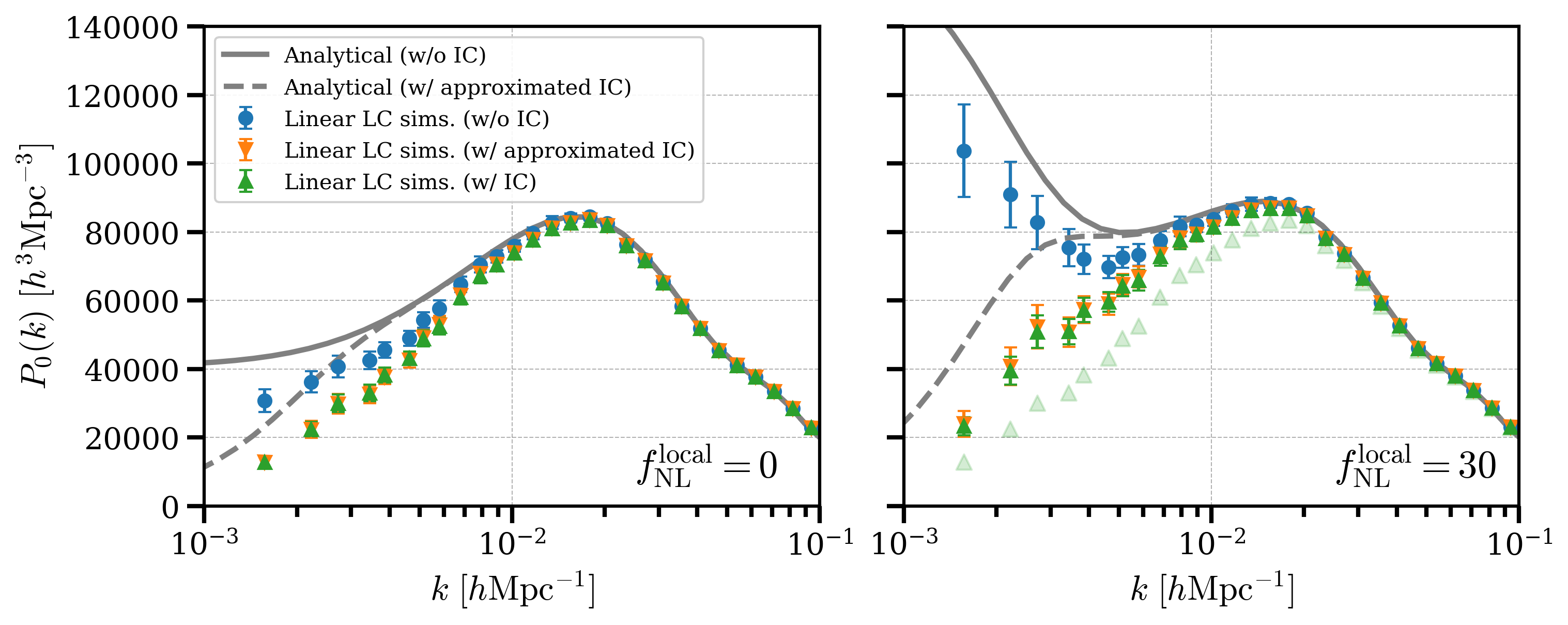}
\includegraphics[width=0.99\textwidth]{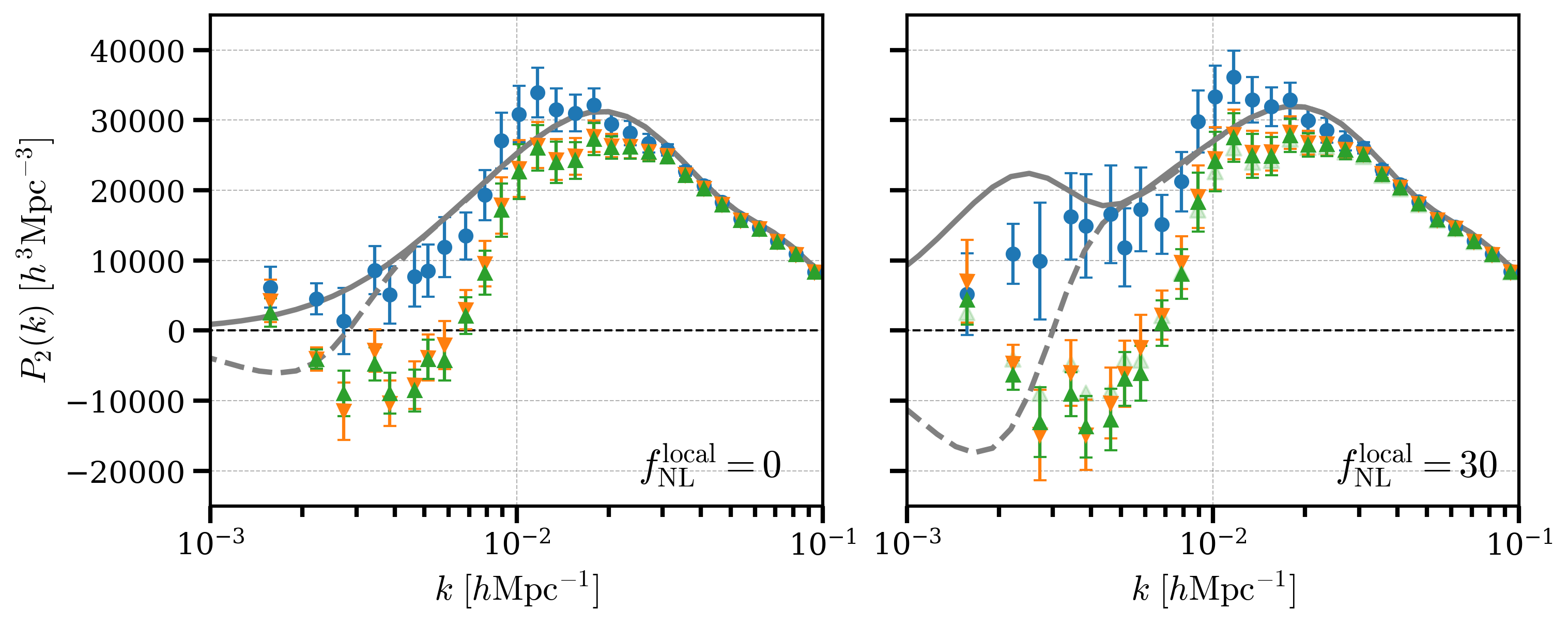}
\includegraphics[width=0.99\textwidth]{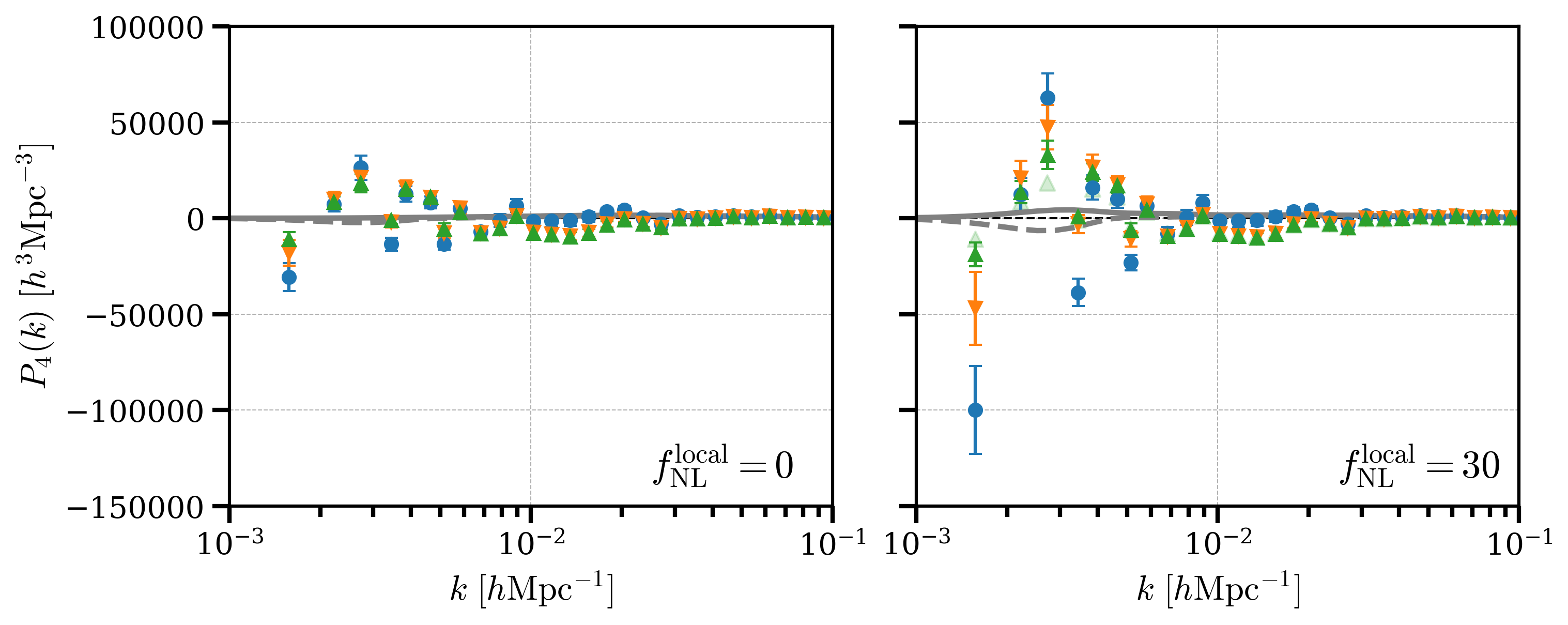}
\caption{Similar to Fig.~\ref{fig:pk0_real}, but showing the monopole, quadrupole, and hexadecapole moments of the redshift-space power spectrum, denoted as 
$P_0$, $P_2$ and $P_4$,  
in the top, middle, and bottom panels, respectively.
For comparison, the light-color symbols in the right panel of each row shows the result for the $f_{\rm NL}=0$ case. 
The inverted triangle symbols show the measured power spectra corresponding to 
the approximated IC expression (Eq.~\ref{eq:approx}), 
where $\overline{\tilde{\Delta}}_b$ is replaced by $\tilde{\Delta}_b(z)$. 
}
\label{fig:pgs_results}
\end{figure}
\begin{figure}
\centering
\includegraphics[width=0.99\textwidth]{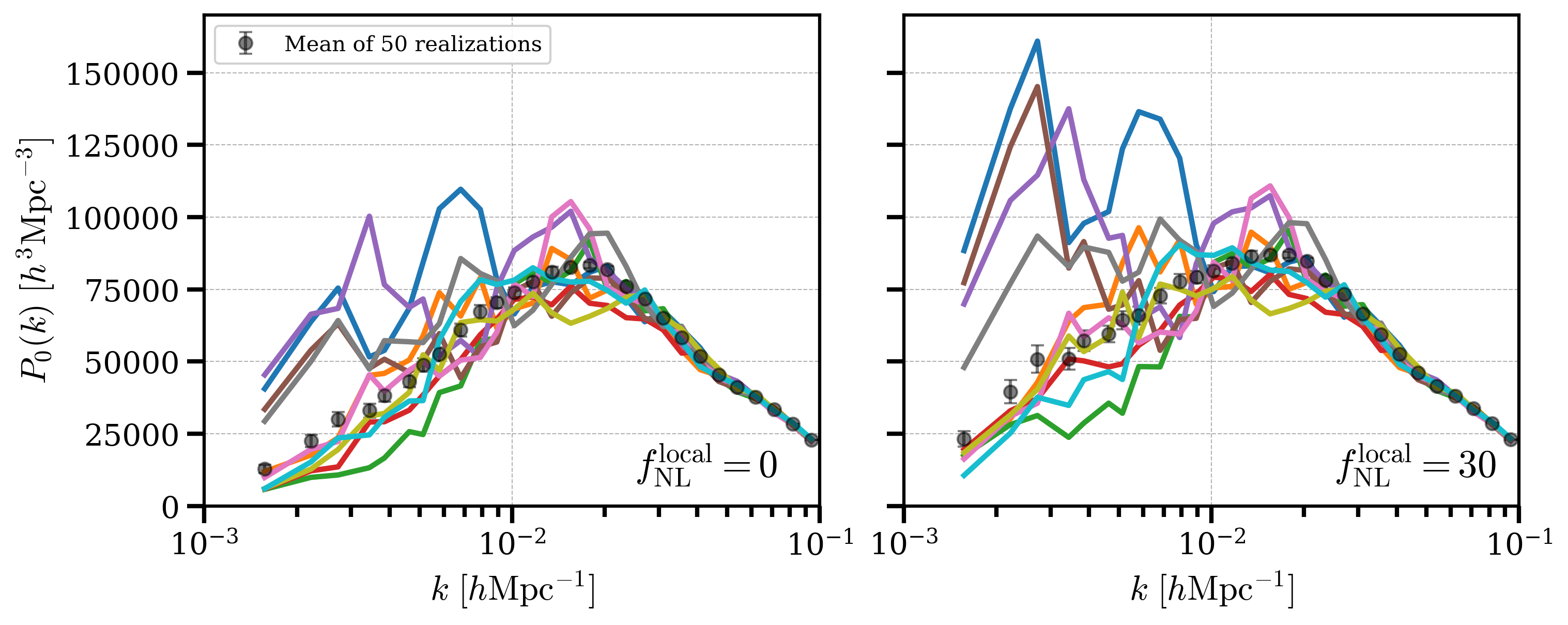}
\includegraphics[width=0.99\textwidth]{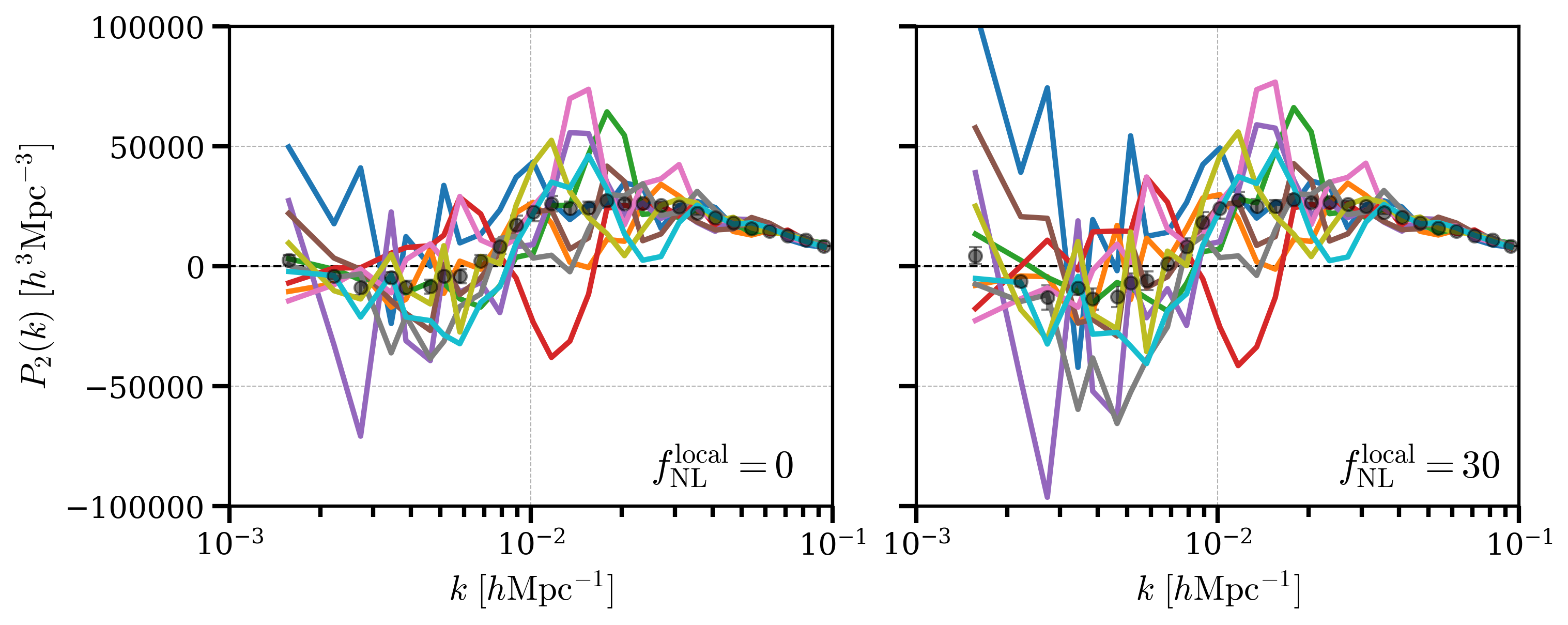}
\caption{Similar to Fig.~\ref{fig:pgs_results}, but each line shows the result for 
each of the 10 light-cone  mock catalogs. The upper and lower panels 
show the results for the monopole and quadrupole moments, respectively.
For comparison, the mean over the 50 realizations, identical to that shown in
Fig.~\ref{fig:pgs_results}, 
is also plotted as black circles in each panel.
}
\label{fig:pgs_sample_variance}
\end{figure}
\begin{figure}
\centering
\includegraphics[width=0.99\textwidth]{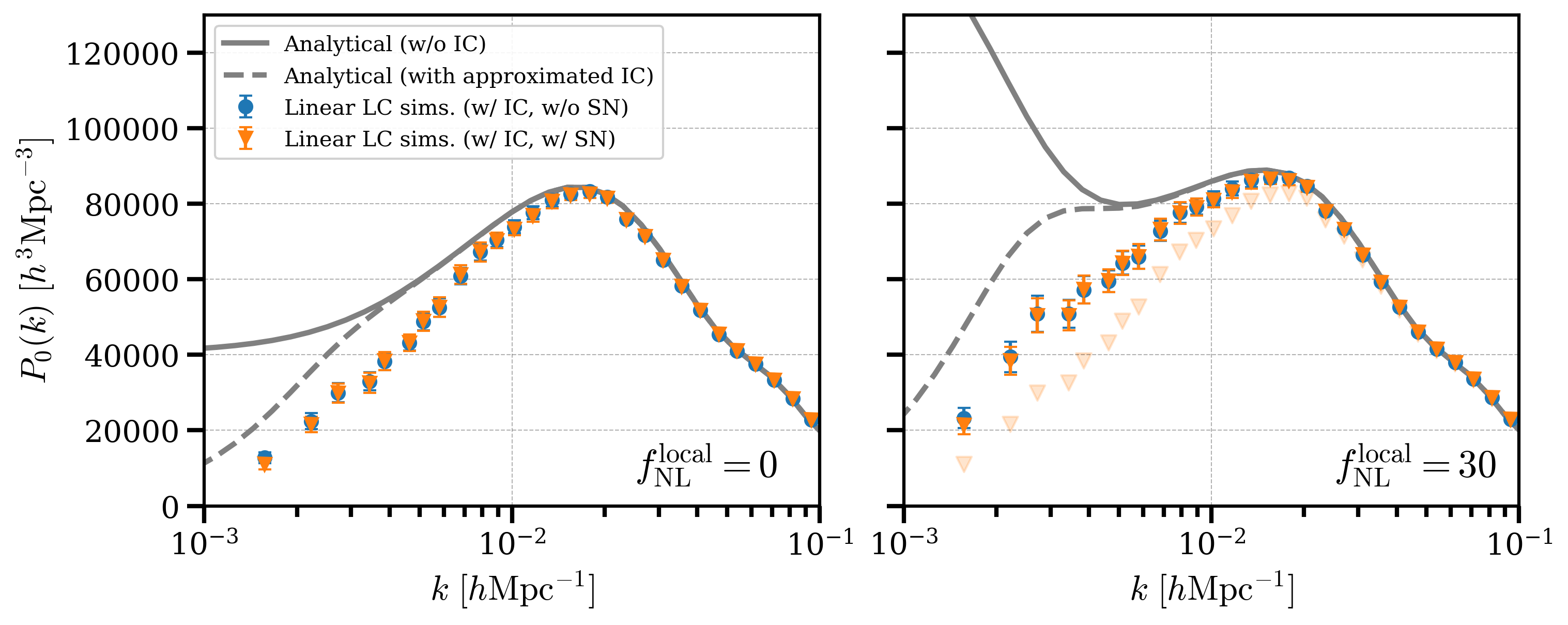}
\includegraphics[width=0.99\textwidth]{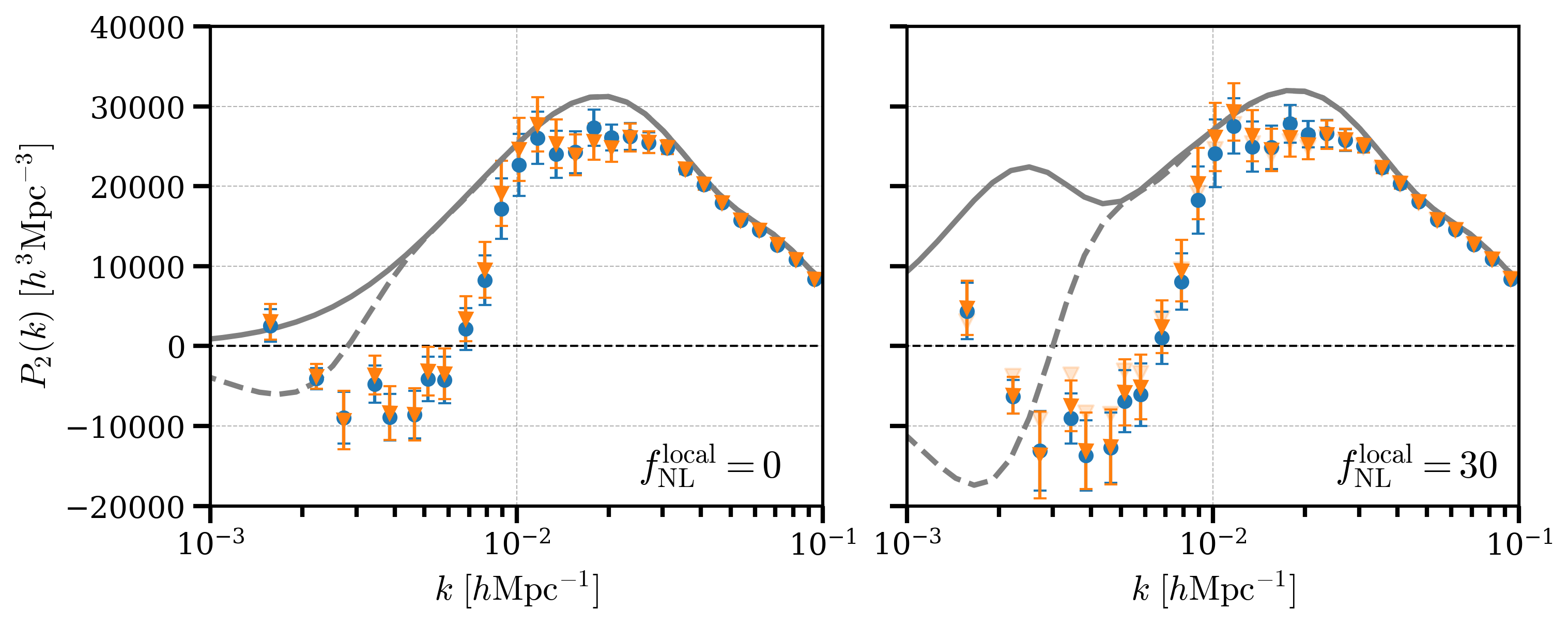}
\caption{The impact of the shot noise on $P_0$ and $P_2$.
The circle symbols are the same as in Fig.~\ref{fig:pgs_results}, but 
the inverted triangle symbols show the results obtained when the shot-noise contribution is included in the light-cone mocks.}
\label{fig:pgs_shotnoise}
\end{figure}
\begin{figure}
\centering
\includegraphics[width=0.6\textwidth]{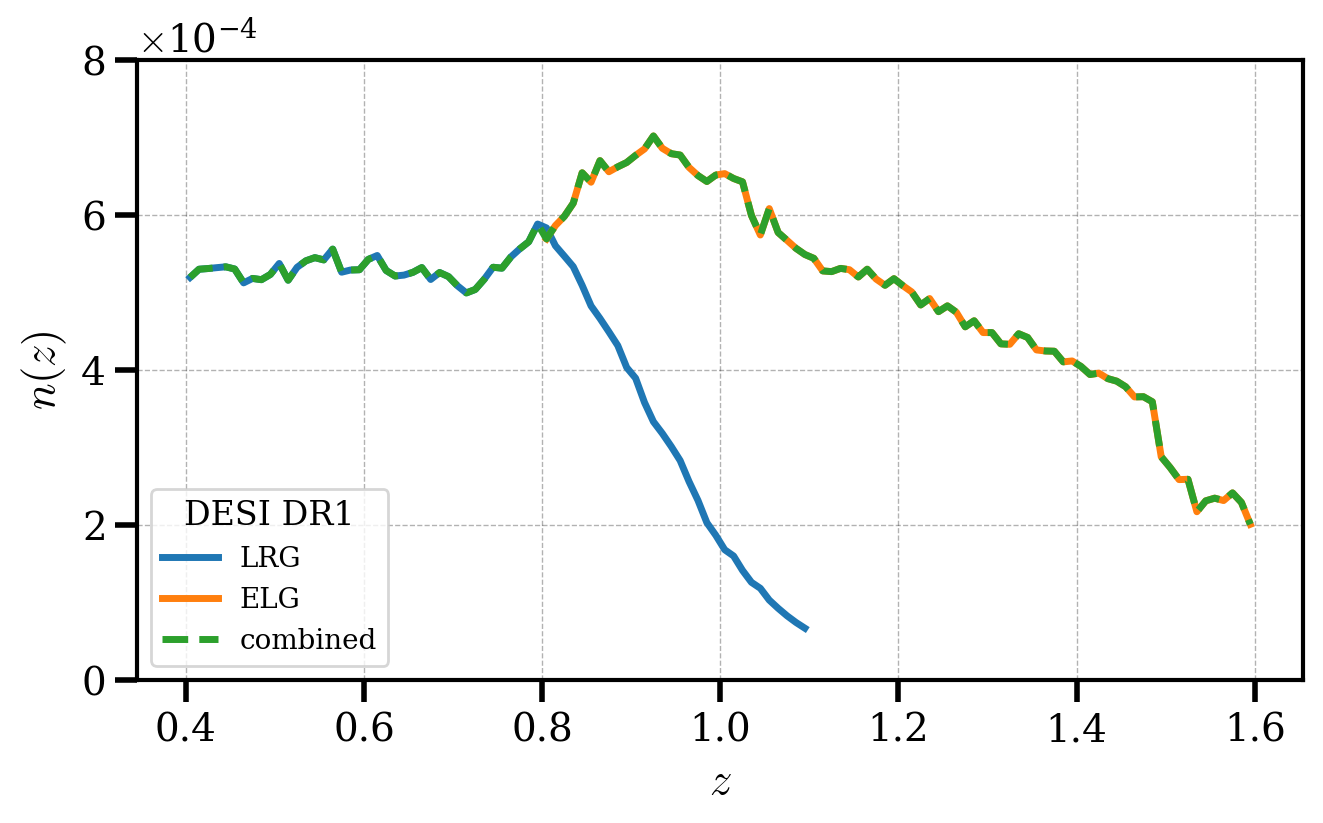}
\caption{The redshift distribution of LRGs and ELGs in the hypothetical DESI survey 
considered in this paper.}
\label{fig:dndz_desi}
\end{figure}
\begin{figure}
\centering
\includegraphics[width=0.99\textwidth]{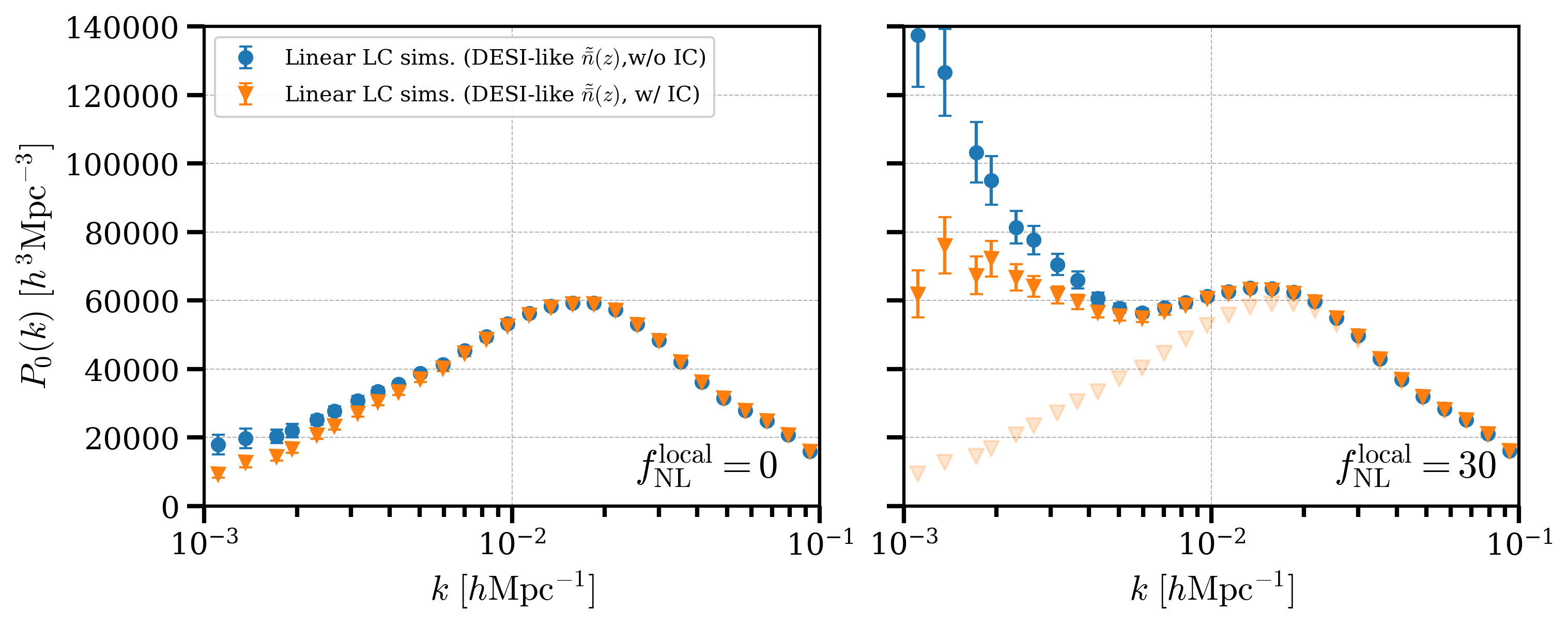}
\caption{The monopole moment of the redshift-space power spectrum measured
from the DESI-like light-cone mock catalogs.}
\label{fig:pgs0_desi}
\end{figure}
\subsection{Real-space power spectrum}
\label{ssec:real-space_pk}

In Fig.~\ref{fig:pk0_real}, 
we show the result for the monopole moment of the power spectrum in {\it real} space, i.e., without the RSD effect.
Here we ignore the shot noise contribution (unless explicitly stated, we will do so
throughout this section).
First, we note that the integral constraint enforces the limit 
$P(k)\rightarrow 0$
as $k\rightarrow 0$.
As a result, it significantly suppresses the PNG-induced excess power, 
$P(k)\propto k^{-4}$, in low $k$ bins. More precisely, the integral constraint removes the apparent divergence of $P(k)$ as $k\rightarrow 0$, implying that the excess power can be observed only down to long-wavelength modes corresponding to the survey-window size, 
$k\sim \mbox{a few}\times O(10^{-3})~h{\rm Mpc}^{-1}$, for a BOSS-like survey. 

For comparison, the solid and dashed lines show the analytical approximation for the window convolution, with and without the IC.
These are computed using Eq.~(\ref{eq:pw_ell_analytic_def}), excluding and including
the second term.  
We adopt  an effective redshift of $z_{\rm eff}=0.462$, defined as the mean 
redshift weighted by $\tilde{\bar{n}}(z)^2$. 
This analytical approximation has been widely used in previous studies
\citep[e.g.,][]{Beutler_2014,FKP_P0,2023PhRvD.108h3533K}.
The analytical prediction without the IC (solid line) agrees well with the light-cone results for both 
$f_{\rm NL}=0$ and 30. 
In contrast, 
the analytical prediction including the IC (dashed line) exhibits a noticeable 
discrepancy from the light-cone mock results at $k\lesssim 0.006~h{\rm Gpc}^{-1}$, corresponding to the size of the BOSS survey. 
This indicates that approximating the IC contribution in Eq.~(\ref{eq:pw_ell_analytic_def}) by the global density contrast $\avrg{(\overline{\tilde{\Delta}}_b)^2}$, instead of the redshift-dependent $\avrg{(\tdeltab(z))^2}$, does not provide an accurate description of the IC.
We also note that, although the galaxy power spectrum for $f_{\rm NL}=30$ 
has an apparent divergence in the limit of $k\rightarrow 0$,
imposing the IC removes the apparent divergence. 
In other words, this indicates 
that a finite-volume survey cannot probe modes with wavelengths larger than the survey scale, and can measure only fluctuations that 
can be decomposed into discrete Fourier modes within the survey volume. 
In Appendix~\ref{sec:box_size}, we compare the results obtained using the light-cone mocks constructed 
from linear simulations with box sizes of
$8$~and 16~$h^{-1}{\rm Gpc}$.

\subsection{Redshift-space power spectrum}
\label{ssec:redshift-space_pk}

Fig.~\ref{fig:pgs_results} shows the main results of this paper.
Here, we used Kaiser formula for the analytical expression of the redshift-space power spectrum (Eq.~\ref{eq:RSD_each_term} without third term).
Compared to Figs.~\ref{fig:pk0_real}, the analytical predictions (solid and dashed lines) cease to be 
accurate at $k\lesssim k_{\rm eq}$, 
which corresponds to the horizon scale at the matter-radiation equality,   
even for the case of $f_{\rm NL}=0$. 
This result therefore implies that it is important to properly account for the survey window effect and other redshift-dependent effects when estimating $k_{\rm eq}$
from the measured galaxy power spectrum.
For the case of $f_{\rm NL}=30$, 
the analytical model performs even worse.
These results clearly demonstrate that the IC, the survey window effect, and light-cone effects 
must all be properly accounted for in order to extract
unbiased cosmological information from the measured multipole moments of the redshift-space power spectrum at $k\lesssim 0.01~h{\rm Mpc}^{-1}$. Here the light-cone effects include the redshift evolution of the underlying density fields and the RSD effect, as well as the redshift distribution of galaxies. We also note that our light-cone mocks include wide-angle effects
\citep{2015MNRAS.447.1789Y}; by applying the same power spectrum estimator to the light-cone mock catalogs as to the actual data, we can directly compare the mock predictions with the measurements. 
The large error bars at the small $k$ bins reflect substantial realization-to-realization variations in $\tdeltab$, as discussed below. 
Furthermore, Fig.~\ref{fig:pgs_results} allows us to assess the validity of the approximation adopted Eq.~\ref{eq:approx} in the presence of PNG, 
a scenario in which the validity of this approximation has not been thoroughly investigated in previous studies. 
In particular, the approximation that the last three IC terms in Eq.~\ref{eq:2pt_def} can be represented by a common value remains valid even in the presence of PNG, whereas the single-redshift approximation, which neglects the redshift dependence of the IC contribution, becomes inaccurate.

In Fig.~\ref{fig:pgs_sample_variance}, we show the results for 10 different 
realizations of the light-cone mocks. Since $\tdeltab$ arises from the super-survey modes and exhibit large realization-to-realization variations (see Fig.~\ref{fig:dndz_deltab}), the measured multipole moments also show substantial variations in small $k$ bins.
Since we adopt the fixed-amplitude approximation (Eq.~\ref{eq:deltam_generation})
when generating the linear fields, the scatters would be larger if the field amplitudes were instead drawn from a Gaussian distribution.

In Fig.~\ref{fig:pgs_shotnoise}, we study how including the shot noise alters the results. 
More specifically, we add a shot-noise contribution to each grid cell, 
estimate the background density including $\tdeltab(z)$ and $\bar{\delta}_{\rm SN}(z)$, 
define the density fluctuation field relative to this background density (see Eq.~\ref{eq:F_def_sim}), 
and then measure the power spectrum.
For the level of shot noise for the BOSS-like survey, as given by $\tilde{\bar{n}}(z)$ (Fig.~\ref{fig:dndz}), 
the results remain almost unchanged. 
Thus, the IC effect in a BOSS-like survey is dominated by sample variance rather than shot noise. Properly accounting for this contribution is therefore essential. 
We also note that,
for surveys with larger shot noise, which can obscure the contributions from super-survey modes, the realization-to-realization scatter becomes significantly larger at small $k$.

\subsection{Forecasts for a DESI-like survey}
\label{ssec:desi}

Finally we discuss the prospects of ongoing and future surveys for constraining PNG.
To this end, we consider a DESI-like survey~\cite{2026AJ....171..285D}, consisting of 
multiple samples of galaxies, namely LRGs and ELGs, spanning redshift range $0.4\le z\le 1.6$. 
Although the actual survey area for DESI is about $15,000$~sq. degrees, we 
adopt the same footprint as the BOSS NGC footprint; the only difference is the redshift range and radial selection function $\tilde{\bar{n}}(z)$.
Fig.~\ref{fig:dndz_desi} shows the redshift distributions of LRG and ELG samples considered in this paper, adopted from DESI DR1~\cite{2026AJ....171..285D}.
Fig.~\ref{fig:pgs0_desi} shows 
the monopole moment of the power spectrum expected from the combined galaxy sample in a DESI-like survey.
Compared to a BOSS-like survey shown in Fig.~\ref{fig:pgs_results}, 
DESI will enable measurements of the power spectrum down to lower $k$, thereby allowing a clearer detection of the PNG-induced excess power at low $k$, if PNG exists.

\section{Conclusion}
\label{sec:conclusion}

In this paper, we have developed a method for constructing light-cone galaxy mock catalogs based on linear-theory simulations that properly incorporate a variety of observational and cosmological effects. 
Since local-type PNG does not alter the matter power spectrum at the level of two-point statistics and affects only the galaxy bias through a characteristic scale-dependent modification on large scales, all relevant fluctuation fields can be generated from a given linear matter power spectrum, $P^L_{\rm m}(k;z)$.
The light-cone mock catalogs account for the survey window function, the redshift evolution of matter and galaxy density fields, RSD,  and the galaxy selection function (the redshift distribution of galaxies). We applied exactly the same procedure for measuring the galaxy power spectrum to each light-cone mock.
More precisely, we 
i) estimate the redshift distribution, i.e., the selection function, from the observed galaxy distribution to define the background density, 
ii) define the 
galaxy density fluctuation field relative to the estimated background density, 
and iii) apply the FKP estimator to measure the multipole moments of galaxy power spectrum from 
each realization.

We paid particular attention to the largest scales accessible in power spectrum measurements from wide-area galaxy surveys.
As a concrete example, we used the light-cone galaxy mocks designed to mimic the BOSS survey, covering the redshift range 
$0.2\le z\le 0.75$ and an area of approximately 8,000~sq. degrees. 
We showed that the effects of the survey-window convolution, discrete Fourier transform 
and the IC on the power spectrum are significant on large scales (i.e., at small $k$), particularly in the presence of PNG
(see Figs.~\ref{fig:pk0_real} and \ref{fig:pgs_results}).
We showed that the analytical approximation commonly used in previous studies breaks down on large scales.
For a BOSS-like survey, failing to properly account for these effects could lead to a biased estimate of 
the matter-radiation equality scale, $k_{\rm eq}$ and the local-type PNG parameter $f_{\rm NL}$, inferred from the measured power spectrum. 
A quantitative assessment of these biases is left for our future work.
We also note that the window-free power-spectrum estimator developed in Ref.~\cite{2021PhRvD.103j3504P} is likely to be affected by super-survey modes. Therefore, the impact of the IC on this estimator should be carefully investigated. We expect that light-cone mocks will provide a useful tool for quantifying these effects.

Our light-cone simulation method could be used for a variety of applications. 
Light-cone mocks can be used not only to extract cosmological information from the measured power spectrum on large, linear scales in an unbiased manner, but also to characterize observational effects on power spectrum measurements.
For example, a random catalog is a fundamental ingredient for quantifying the galaxy selection function and is indispensable for measurements of clustering statistics. 
However, the construction of a random catalog necessarily relies on the observed galaxy distribution itself, and it is therefore unavoidable that the distribution is affected by large-scale density fluctuations, i.e. super-survey modes (see Fig.~\ref{fig:dndz_deltab}). 
Light-cone mock catalogs are thus useful for quantifying the importance of the accuracy of random-catalog construction, as well as the impact of its uncertainties on power spectrum measurements. In particular, they provide a powerful tool for 
quantifying the impact of PNG and various systematic effects on measurements of the large-scale power spectrum~\citep[see][for a similar discussion]{2012MNRAS.424..564R,2017MNRAS.464.1168R,Hand.Li.ea2017jul,Hand.Li.ea2017jul}.
Another possible application is to extend the method developed in this paper to incorporate more general forms of PNG or large-scale isocurvature modes.
Rather than relying on analytical modeling of these effects on the power spectrum, light-cone mocks can directly model the measured power spectrum and bispectrum.
We hope that the methods developed in this paper will facilitate these applications in preparation for upcoming wide-area galaxy surveys.

\acknowledgments
We would like to thank Kaz~Akitsu and Steven~Chen for very useful discussion.  
This work was supported in part by
JSPS KAKENHI Grant Number 24H00215, 26H00401,
26H00404
and by World Premier International Research Center Initiative (WPI Initiative), MEXT, Japan.

\appendix

\section{Dependence of light-cone mocks on linear-simulation box size}
\label{sec:box_size}

\begin{figure}
\centering
\includegraphics[width=0.99\textwidth]{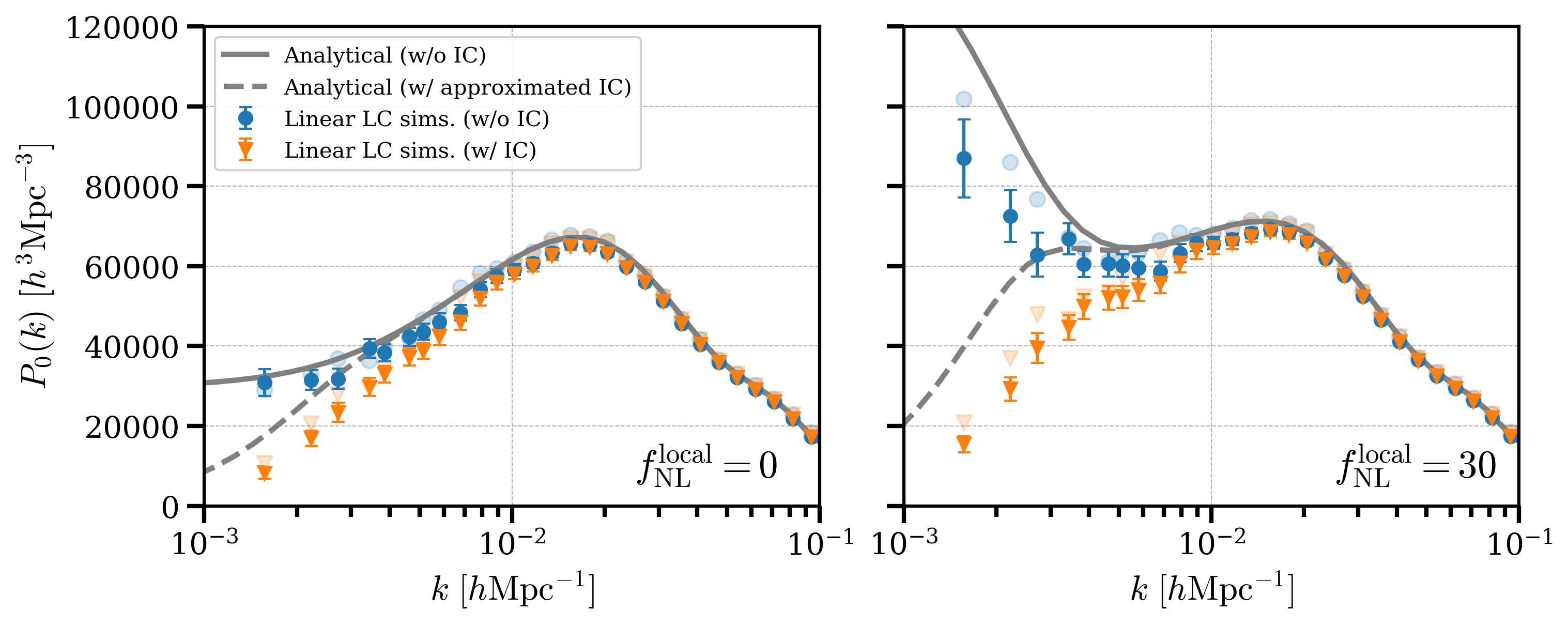}
\includegraphics[width=0.99\textwidth]{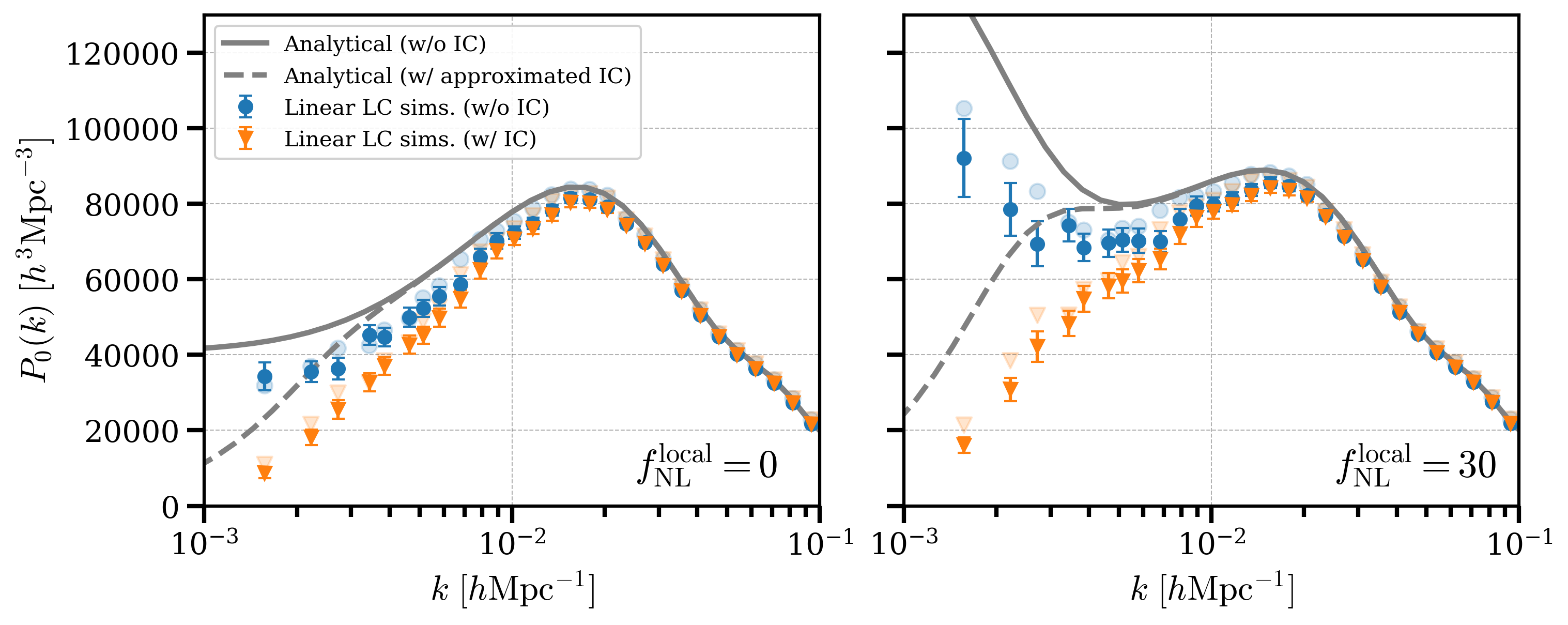}
\caption{Comparison of the monopole moments of power spectra with and without the IC, measured from the light-cone mock catalogs that 
are constructed from linear-theory simulations with our fiducial box size of $16~h^{-1}{\rm Gpc}$ (light-colored markers) and a smaller box size of 
$8~h^{-1}{\rm Gpc}$ (dark-colored makers with error bars), respectively. 
The upper and lower two panels show the real-and redshift-space power spectra, respectively, 
while the left and right panels in each row correspond to $f_{\rm NL}=0$ and $30$, respectively.
The results for the $16~h^{-1}{\rm Gpc}$ box (light-colored markers) are identical to those in Figs.~\ref{fig:pk0_real} and \ref{fig:pgs_results}.
The solid line shows the result for the analytical prediction for the window-convolved power spectrum without the IC
(the first term of Eq.~\ref{eq:pw_ell_analytic_def}). 
The solid lines for the real-space power spectrum in the upper panels
exhibit better agreement with the light-cone results
from $16~h^{-1}{\rm Gpc}$ box  simulation, compared with the results from the $8~h^{-1}{\rm Gpc}$ simulation.  
For the other cases, however, the analytical prediction fails to reproduce the light-cone results.
}
\label{fig:8_vs_16Gpc}
\end{figure}
In this section, we study the effect of linear-simulation box size on light-cone mock results. 
Since the main focus of this paper is the impact of super-survey modes on galaxy power-spectrum measurements, it is important to use linear-theory simulations with box sizes larger than the survey window, such as that of the BOSS NGC footprint.
Throughput this paper we have used the simulations with a box size of $16~h^{-1}{\rm Gpc}$ as our default setup. 
In Fig.~\ref{fig:8_vs_16Gpc} compares the results obtained from linear simulations with box sizes of 8 and 16$~h^{-1}{\rm Gpc}$, respectively. 
The power spectra change slightly at $k\lesssim k_{\rm eq}$ 
due to the additional long-wavelength modes included in the input linear power spectrum of the
16$~h^{-1}{\rm Gpc}$ box simulation, for both $f_{\rm NL}=0$ and $30$ cases.
Since linear-theory simulations are computationally inexpensive, this effect can be straightforwardly quantified and corrected for.
The solid and dashed lines in each panel show the analytical predictions for the power spectrum without and with the IC, respectively (Eq.~\ref{eq:pw_ell_analytic_def} excluding and including the second term).
Only for the real-space power spectrum without the IC
does the analytical prediction exhibit better agreement with 
the light-cone results from the 
$16~h^{-1}{\rm Gpc}$ box simulation than with those from the $8~h^{-1}{\rm Gpc}$ box simulation. 
This implies that the approximations used in Eq.~(\ref{eq:pw_ell_analytic_def}), such as the use of a single effective redshift, are valid
for the real-space power spectrum with the IC. 
However, for the other cases, including the real-space power spectrum with the IC, the analytical predictions lose accuracy in the low-$k$ regime. This suggests that the IC induced by $\tdeltab(z)$ in each redshift shell, as well as other redshift-dependent effects such as RSD,  
are not properly captured by the analytical model.

\bibliographystyle{JHEP}
\bibliography{refs}
\end{document}